\pdfoutput=1
\documentclass[11pt,reqno,preprint]{article}
\usepackage{jheppub,epsfig,amssymb,amsmath,mathrsfs,hyperref,graphicx}
\usepackage{multirow,bbm}
\usepackage[makeroom]{cancel}
\usepackage{caption}
\usepackage{subcaption}
\usepackage{ulem}
\usepackage[dvipsnames]{xcolor}

\def\be{\begin{equation}}
\def\ee{\end{equation}}
\def\ba{\begin{eqnarray}}
\def\ea{\end{eqnarray}}
\newcommand{\bea}{\begin{eqnarray}}
\newcommand{\eea}{\end{eqnarray}}
\newcommand{\nn}{\nonumber}

\def\Li{\textrm{Li}}

\def\e{\epsilon}

\def\Eqn#1{Equation~(\ref{#1})}
\def\eqns#1#2{eqs.~(\ref{#1}) and~(\ref{#2})}

\def\Eqn#1{Equation~(\ref{#1})}
\def\eqn#1{eq.~(\ref{#1})}
\def\eqns#1#2{eqs.~(\ref{#1}) and~(\ref{#2})}

\def\emph#1{{\it #1}}

\newcommand{\fwboxL}[2]{\text{\makebox[#1][l]{$#2$}}}

\def\EE{\mathcal{E}}

\def\Gcusp{\Gamma_{\rm cusp}}

\def\trphi#1{{\rm Tr}\,\phi^{#1}}

\newcommand{\cC}{\begin{cal}C\end{cal}}

\newcommand{\cH}{\begin{cal}H\end{cal}}

\newcommand{\cL}{\begin{cal}L\end{cal}}
\newcommand{\cM}{\begin{cal}M\end{cal}}

\newcommand{\cW}{\begin{cal}W\end{cal}}

\def\blue#1{{\color{blue}#1}}
\def\green#1{{\color{Green}#1}}
\def\red#1{{\color{red}#1}}

\title{The Three-Point Form Factor of $\trphi3$ to Six Loops}

\author{Benjamin Basso$^{1,3}$,}
\author{Lance~J.~Dixon$^{2,3}$} 
\author{and Alexander~G.~Tumanov$^{1,4}$}

\affiliation{$^{1}$ Laboratoire de physique de l'Ecole normale sup\'erieure,
ENS, Universit\'e PSL, CNRS \\
Sorbonne Universit\'e, Universit\'e Paris-Diderot, Sorbonne Paris Cit\'e,
75005 Paris, France}

\affiliation{$^{2}$ SLAC National Accelerator Laboratory,
Stanford University, Stanford, CA 94309, USA}

\affiliation{$^{3}$ Simons Center for Geometry and Physics,
SUNY, Stony Brook, NY 11794, USA}

\affiliation{$^{4}$ Institut Philippe Meyer, Ecole normale sup\'erieure, 75005 Paris, France}

\abstract{We study the three-point form factor of the length-three half-BPS operator ($\trphi3$) in planar $\mathcal{N}=4$ Super-Yang-Mills theory, using analyticity and integrability methods. We find that the functions describing the form factor in perturbation theory live in the same restrictive space of multiple polylogarithms as the one describing the form factor of the stress-tensor operator ($\trphi2$). Furthermore, we find that the leading-order data in the collinear limit provided by the form factor operator product expansion (FFOPE) is enough to fix the form factor uniquely, at least through six loops. We perform various tests of our results using the subleading FFOPE corrections. We also analyze the form factor in the Regge limit where two Mandelstam invariants are large; we obtain a compact representation for the form factor in this limit which is valid to all orders in the coupling.}

\emailAdd{benjamin.basso@phys.ens.fr}
\emailAdd{lance@slac.stanford.edu}
\emailAdd{alexander.tumanov@phys.ens.fr}

\begin{document}
\hypersetup{pageanchor=false}
\maketitle
\hypersetup{pageanchor=true}


\newcommand\scalemath[2]{\scalebox{#1}{\mbox{\ensuremath{\displaystyle #2}}}}

\section{Introduction}
\label{sec:intro}
The idea that scattering amplitudes, as functions of their kinematic variables, may be determined from a few basic principles dates back to the early days of quantum field theory. It was first explored at the time of the analytic S-matrix program, in the early 60s, when it became clear that general concepts, such as unitarity and causality, may be used to constrain the form of scattering amplitudes~\cite{ELOP}. In recent years, a new incarnation of this idea has emerged in the study of loop amplitudes in the planar limit of the maximally supersymmetric Yang-Mills theory ($\mathcal{N}=4$ SYM). There, it became apparent that scattering amplitudes belong to a restricted space of transcendental functions, which encodes in a minimal way the physical requirements of the theory. A prime example is the maximally-helicity violating (MHV) six-gluon amplitudes, which were argued to belong to a small space of multiple polylogarithms (MPLs), at each loop order~\cite{Goncharov:2010jf,Dixon:2011nj}. This crucial observation enabled the development of a powerful amplitude bootstrap program, which constructs amplitudes directly by leveraging the knowledge of the mathematical space of functions they inhabit. Over the past decade, this approach has achieved remarkable success in calculating a variety of higher-loop six- and seven-point amplitudes in planar $\mathcal{N}=4$ SYM~\cite{Dixon:2011pw,Dixon:2011nj,Dixon:2013eka,Dixon:2014voa,Dixon:2014iba,Drummond:2014ffa,Dixon:2015iva,Caron-Huot:2016owq,Dixon:2016apl,Dixon:2016nkn,Drummond:2018caf,Caron-Huot:2019vjl,Dixon:2023kop}.
Another method for computing higher-loop amplitudes in planar $\mathcal{N}=4$ SYM relies on the anomalous action of the dual conformal symmetry generator, the ${\overline Q}$ equation~\cite{Caron-Huot:2011dec,Bullimore:2011kg}.  The ${\overline Q}$ method has also seen great success in computing amplitudes at two or three loops with eight or more external legs~\cite{Caron-Huot:2011dec,Caron-Huot:2011zgw,Caron-Huot:2013vda,He:2019jee,He:2020vob,Li:2021bwg,He:2022ujv}.

The application of the amplitude bootstrap method to form factors of local operators in planar $\mathcal{N}=4$ SYM is more recent. 
It was initiated in refs.~\cite{Dixon:2020bbt,Dixon:2022rse} for the MHV form factors of the stress-energy tensor. Progress in this area was made possible by the availability of invaluable boundary data in the collinear limit, which helped design the relevant spaces of functions for the form factors. This boundary data was generated through an exact integrability-based formalism known as the Form Factor Operator Product Expansion (FFOPE) method~\cite{Sever:2020jjx,Sever:2021nsq,Sever:2021xga,Alday:2010ku}. It originates from the existence of a dual description of form factors in terms of null periodic Wilson loops~\cite{Alday:2007hr,Alday:2007he,Maldacena:2010kp,Brandhuber:2010ad,Bork:2014eqa,Sever:2020jjx}.
The FFOPE method also incorporates elements from the OPE method developed earlier for scattering amplitudes (closed null Wilson loops)~\cite{Basso:2013vsa,Basso:2013aha,Basso:2014koa,Basso:2014hfa,Basso:2015rta,Basso:2015uxa,Belitsky:2014sla,Belitsky:2014lta,Belitsky:2016vyq}; the OPE method similarly supplied critical boundary data for the amplitude bootstrap.

Recently, the Wilson-loop FFOPE approach was extended to the form factors of all half-BPS operators $\textrm{Tr}\,\phi^k$ with $k\geq2$, where $\phi$ denotes an arbitrary complex scalar field~\cite{Basso:2023bwv}.  Such operators belong to short representations of the superconformal group, with a conformal dimension $k$ that is protected by supersymmetry. The simplest of these multiplets with $k=2$ coincides with the stress-tensor multiplet of $\mathcal{N}=4$ SYM, which contains both $\textrm{Tr}\,\phi^2$ and the stress-tensor operator mentioned before. The higher multiplets with dimension $k\geq 3$ are its simplest generalizations, which we also expect to be amenable to the bootstrap approach with the help of the integrability description.

This paper focuses on bootstrapping the three-point form factor of the operator $\textrm{Tr}\,\phi^3$. This form factor was studied previously, analytically at two loops~\cite{Brandhuber:2014ica} and numerically at three loops~\cite{Lin:2021kht,Lin:2021qol}. The latter results are based on numerical integration of integrands constructed using generalized unitarity and color-kinematics duality.
(Higher-point form factors have been obtained using similar methods~\cite{Guo:2021bym,Guo:2024bsd}, and also bootstrapping with a knowledge of all the master integrals.)An analytic three-loop calculation was also performed very recently~\cite{Johannes}. In this work, we extend the state-of-the-art for this observable through six loops using the amplitude/form-factor bootstrap technique.

Similarly to the three-point form factor of the stress-tensor operator~\cite{Dixon:2020bbt,Dixon:2022rse}, the three-point form factor of $\textrm{Tr}\,\phi^3$ can be embedded into a space of multiple polylogarithms at any order in the loop expansion. These functions are defined recursively as iterated integrals over logarithmic
kernels~\cite{%
  Chen,G91b,Goncharov:1998kja,Remiddi:1999ew,Borwein:1999js,Moch:2001zr},
\begin{equation} \label{eq:G_func_def}
G_{a_1,\dots, a_n}(z) = \int_0^z \frac{dt}{t-a_1}\,G_{a_2,\dots, a_n}(t)\ , \qquad G_{\fwboxL{27pt}{{\underbrace{0,\dots,0}_{p}}}}(z) = \frac{\ln^p z}{p!} \ .
\end{equation}
The usual polylogarithms form a particular subclass of these functions,
\begin{equation}
\Li_n(z) = -\,G_{\fwboxL{34pt}{{\underbrace{0,\dots,0}_{n-1},1}}}(z) \ .
\end{equation}
Harmonic polylogarithms (HPLs)~\cite{Remiddi:1999ew} $H_{\vec{a}}(z)$ with $a_i \in \{0,1,-1\}$ can also be expressed in terms of these functions, using 
\begin{equation}
H_{\vec{a}}(z)=(-1)^p\,G_{\vec{a}}(z)\ ,
\end{equation}
where $p$ is the number of 1's in the string $\vec{a}$.
We find that, rather remarkably, the three-point form factor of $\textrm{Tr}\,\phi^3$ fits into the same space of polylogarithmic functions $\mathcal{C}$ as its $\textrm{Tr}\,\phi^2$ counterpart~\cite{Dixon:2022rse}. It does, however, obey a different set of multi-final entry conditions. Together with the boundary FFOPE results, they are sufficient to bootstrap the form factor through six loops.

Lastly, we will study a particular ``Regge" limit, where two of the three Mandelstam invariants are sent to infinity. Interest in this limit arises from the observation that it admits an all-order description that extends beyond the collinear limit. Furthermore, it bears striking similarities with the multi-particle factorization limit of the Next-to-MHV (NMHV) six-point amplitudes, which was studied through higher loops in refs.~\cite{Dixon:2014iba,Dixon:2015iva}, and with the self-crossing limit of six-point amplitudes~\cite{Dixon:2016epj,Caron-Huot:2019vjl}. To be more precise, we will find that the form factor is singular in the Regge limit and exhibits logarithmic singularities which may be resummed to all loops using the FFOPE method.

The paper is organized as follows. In section~\ref{sec:FFspaceReview}, we review the form factor bootstrap procedure, the class of form factors to which it will be applied, and the properties of the associated function space. Section~\ref{sec:Constraints} focuses on the classification and imposition of the constraints. The results are presented and analyzed in section~\ref{sec:Properties}. Finally, in section~\ref{sec:FFOPE}, we demonstrate how the leading-order boundary data is obtained using the FFOPE approach and leverage it to determine the analytic behavior of the form factor in the Regge kinematics. The appendices include the computation of the subleading FFOPE correction, along with a presentation of all the necessary building blocks for its construction. 

Ancillary files attached to the paper include: {\tt WL\_FFOPE.txt}, containing the closed-form leading and subleading FFOPE results, and {\tt phi3FFmultifinalentry.txt}, detailing the multi-final entry conditions obeyed by the form factor.
{\tt EE33G.txt} provides the form factor results as $G$ functions through 5 loops (at 6 loops they are too lengthy), while {\tt phi3FFsymbols.txt} provides the symbols through 6 loops.
We also provide ancillary files associated with a ``sewing matrix'' representation of the symbols of the $\trphi2$ and $\trphi3$ form factors, {\tt Cfrontspace.txt}, {\tt phi2backspace.txt}, {\tt phi3backspace.txt}, {\tt phi2SewMatrix.txt}, and {\tt phi3SewMatrix.txt}.


\section{Review of form-factor bootstrap}
\label{sec:FFspaceReview}

\subsection{Super form factors}\label{section:super_FF_def}
In $\mathcal{N}=4$ Super-Yang-Mills theory, half-BPS superfields $\mathcal{T}_{k}(x,\theta)$ are defined by acting on the chiral primary operators $\trphi{k}(x)$ with chiral supercharges $Q^{\alpha A}$~\cite{Penante:2014sza,Eden:2011yp,Eden:2011ku},
\begin{equation}\label{eq:calT-k}
    \mathcal{T}_{k}(x, \theta) = e^{\theta_{\alpha A} Q^{\alpha A}}\cdot \textrm{Tr} \, \phi(x)^k = \textrm{Tr} \, \phi(x)^k + \theta_{\alpha A} \left[Q^{\alpha A}, \textrm{Tr} \, \phi(x)^k\right] + \ldots \, .
\end{equation}
where $\phi$ denotes a complex scalar field. The Grassmann variables $\theta_{\alpha A}$ parametrize the multiplet, while their anti-chiral counterparts $\bar{\theta}_{\dot{\alpha} A}$ are set to zero. Note that due to the fact that the chiral primary operator $\textrm{Tr} \, \phi(x)^k$ is protected, it is annihilated by half of the supercharges $Q^{\alpha A}$, which implies that only half of the $\theta_{\alpha A}$ variables are present in eq.~(\ref{eq:calT-k}). We will denote the $\theta$-variables associated with the generators that annihilate the operator as $\theta^-_{\alpha\,a}$, $a=1,2$, while the ones that remain intact will be called $\theta^+_{\alpha\,a'}$, $a'=3,4$.\par
The super form factors of such operators are given by the super Fourier transform of the corresponding matrix element~\cite{Brandhuber:2011tv,Penante:2014sza},
\begin{equation}\label{eq:Fourier-Fkn}
   \mathcal{F}_{k,n}\left(1,\ldots,n;q,\gamma\right) = \int d^{4}x d^{4}\theta \, e^{-iq_{\mu} x^{\mu} - \gamma^{+\alpha a'}\theta^+_{\alpha a'}} \langle1,\ldots,n|\mathcal{T}_{k}(x, \theta^+)|0\rangle\ ,
\end{equation}
where the Fourier parameters $q$ and $\gamma^+$ are the momentum and supermomentum carried by the operator, respectively. The supermomentum $\gamma^-$, for the supercharges that annihilate the chiral primary operator, is naturally equal to zero.

Similarly to the operator, each of the external states, labeled by $i=1,\ldots,n$ in \eqn{eq:Fourier-Fkn}, and carrying light-like momenta $p_i = \lambda_i\tilde{\lambda}_i$, likewise belongs to a supermultiplet~\cite{Nair:1988bq},
\begin{equation}
\Phi(p,\tilde{\eta}) = g^+(p) + \tilde{\eta}^A\psi_A(p)+\frac{\tilde{\eta}^A\tilde{\eta}^B}{2!}\,\phi_{AB}(p)+\epsilon_{ABCD}\frac{\tilde{\eta}^A\tilde{\eta}^B\tilde{\eta}^C}{3!}\,\bar{\psi}^D(p)+\tilde{\eta}^1\tilde{\eta}^2\tilde{\eta}^3\tilde{\eta}^4\,g^-(p)\ ,
\end{equation} 
where $\{g^+,\psi_A,\phi_{AB},\bar{\psi}^D,g^-\}$ label the type of particle being scattered, and $\tilde{\eta}^{A}$, with $A=1,2,3,4$, are bookkeeping Grassmann parameters.

At tree level, the MHV form factors of the stress-tensor multiplet $k=2$ take the simple form~\cite{Brandhuber:2010ad,Brandhuber:2011tv}, resembling the Parke-Taylor-Nair formula~\cite{Parke:1986gb,Nair:1988bq},
\begin{align}
\mathcal{F}^{\rm MHV,\,tree}_{2,n}\left(1,\ldots,n;q,\gamma^{+}\right) = \frac{\delta^{(4)}\left(q-\sum\limits_{i=1}^n\lambda_i\tilde{\lambda}_i\right)\delta^{(4)}\left(\gamma^+-\sum\limits_{i=1}^n\lambda_i\tilde{\eta}^+_{i}\right)\delta^{(4)}\left(\sum\limits_{i=1}^n\lambda_i\tilde{\eta}^-_{i}\right)}{\langle12\rangle\,\langle23\rangle\,\ldots\,\langle n1\rangle}\ .
\end{align}
The form factors of general half-BPS multiplets $\mathcal{T}_k$ involve additional helicity-carrying dressing factors. For instance, in the $\mathcal{T}_{3}$ case, one finds~\cite{Penante:2014sza}
\begin{align}\label{F3n}
\mathcal{F}^{\rm MHV,\,tree}_{3,n} &= -\,\mathcal{F}^{\rm MHV,\,tree}_{2,n}\sum\limits_{1\leqslant i<j \leqslant n}\langle ij\rangle\,\tilde\eta^-_{i}\cdot\tilde\eta^-_{j}\ ,
\end{align}
where $\langle ij\rangle = \epsilon_{\alpha\beta}\lambda^{\alpha}_{i}\lambda^{\beta}_{j}$ is the boost invariant product of spinor helicity variables, $\tilde\eta^-_{i}\cdot\tilde\eta^-_{j} = \frac{1}{2}\epsilon_{ab}\tilde{\eta}_{i}^{a}\tilde{\eta}_{j}^{b}$, and $\epsilon$ is the Levi-Civita tensor with $\epsilon_{12} = -\,\epsilon_{21} = 1$.  

The so-called minimal form factor case corresponds to the number of external particles $n$ being equal to the number of scalars in the operator $k$. At any loop order, these form factors $\mathcal{F}_{n,n}$ are proportional to their tree-level contributions, in the sense that the dependence on the fermionic variables can be fully factored out, leaving one with an essentially bosonic object. At one loop, with loop-expansion parameter $g^2 \equiv N g_{\rm YM}^2/(16\pi^2) = \lambda/(16\pi^2)$, the result is a sum of $n$ scalar triangle integrals~\cite{vanNeerven:1985ja,Brandhuber:2010ad,Brandhuber:2014ica},
\begin{align}\label{F3n1l}
\mathcal{F}^{\text{MHV,\,\rm{1-loop}}}_{n,n} &= \,\mathcal{F}^{\text{MHV,\,\rm{tree}}}_{n,n}\,M_{n,n}(\e)\ ,
\end{align}
with
\begin{align}
M_{n,n}(\e)\ =\
-\,\frac{1}{\e^2} \sum_{i=1}^{n} \left(\frac{4\pi e^{-\gamma}\mu^2}{-s_{i,i+1}}\right)^\e
+ \frac{n}{2} \, \zeta_2 + \mathcal{O}(\epsilon) \, ,
\end{align}
where $s_{i,i+1} = (p_{i}+p_{i+1})^2$, $\gamma$ is the Euler-Mascheroni constant, and $\epsilon = \frac{1}{2}(4-D)$ with $D$ the spacetime dimension.

At higher loops, the infrared-divergent part of the form factor exponentiates (see e.g.~refs.~\cite{Sudakov:1954sw,vanNeerven:1985ja,Bern:2005iz,Brandhuber:2012vm,Brandhuber:2014ica}).  The rest is captured by a finite, $\e$-independent remainder function $R_{n,n}$, leading to the following all-loop expression
\begin{equation}
\frac{{\cal F}^{\rm MHV}_{n,n}}{{\cal F}_{n,n}^{\rm MHV,\,tree}}\ =\
\exp\,\biggl\{ \sum_{L=1}^\infty g^{2L}
\biggl[ \Bigl( \frac{\Gcusp^{(L)}}{4}\, +\, {\cal O}(\e) \Bigr)
  M_{n,n}(L\e)\, +\, C_{n,n}^{(L)}
  \, +\, R_{n,n}^{(L)} \biggr]
  \biggr\}\,.
\label{RemainderFnDef}  
\end{equation}
Here, the terms in brackets are the same as those entering the Bern-Dixon-Smirnov (BDS) ansatz for amplitudes~\cite{Bern:2005iz}, $C_{n,n}^{(L)}$ are some scheme-dependent constants, and $\Gcusp^{(L)}$ denotes the $L$-loop coefficient of the cusp anomalous dimension~\cite{Korchemsky:1987wg}. In planar ${\cal N}=4$ SYM, $\Gcusp$ is known to all loop orders~\cite{Beisert:2006ez},
\begin{eqnarray}
\Gcusp(g) 
&=& \sum_{L=1}^\infty g^{2L} \, \Gcusp^{(L)} 
\nonumber\\
&=& 4\,g^{2} - 8\,\zeta_{2}\,g^{4} + 88\,\zeta_{4}\,g^{6} - \left(876\,\zeta_{6} + 32\,\zeta_{3}^{2}\right) g^{8} + \mathcal{O}(g^{10})\, .
\label{cuspeq}
\end{eqnarray}
Through two loops, $R_{n,n}$ is known explicitly for any $n$~\cite{Brandhuber:2014ica}. 

The constant $C_{n,n} = \sum_{L=1}^{\infty} g^{2L} C_{n,n}^{(L)}$ depends on the normalization of the remainder function $R_{n,n}$. For the Sudakov form factor, there is no non-trivial remainder function, $R_{2,2} = 0$, and $C_{2,2}$ is defined as the leftover constant in the BDS-subtracted form factor. Explicit calculations in the gauge theory, taking into account also a three-loop constant in the BDS ansatz, have determined it through three loops~\cite{Brandhuber:2012vm,Guan:2023gsz,Chen:2023hmk}:
\begin{equation}\label{eq:C22}
C_{2,2} = 4\,\zeta_{4}\,g^{4} 
+ \biggl(-\,\frac{181}{3}\,\zeta_6 + 16\,\zeta_{3}^2 \biggr)\,g^{6} + \mathcal{O}(g^8)\, .
\end{equation}
For the higher minimal form factors, the remainder function $R_{n,n}$ is non-trivial and the definition of the constant $C_{n,n}$ is more arbitrary. Throughout this paper, we will find convenient to work with $C_{n,n} = C_{2,2}$, to all loops. Specifically, we choose
\begin{equation}\label{eq:C33isC22}
C_{3,3} = C_{2,2} \,.
\end{equation}
Note that this choice may differ from others used in the literature. In particular, it differs from the choice made in refs.~\cite{Brandhuber:2014ica,Lin:2021qol} for $\trphi3$, where the constant term in the BDS-subtracted three-point form factor was absorbed in the definition of the remainder function. As a result, our remainder function for $n=3$ departs slightly from the one constructed in these references. Namely, one has
\begin{equation}
R_{3,3} = R_{3,3}^{\textrm{\cite{Brandhuber:2014ica}}} -C_{2,2}\, ,
\end{equation}
where $C_{2,2}$ is the constant in eq.~\eqref{eq:C22}.

The focus of this paper is the three-point form factor, $n=3$, where $R_{3,3} = R_{3,3}(u, v, w)$ is a function of the kinematic invariants~\cite{Brandhuber:2014ica,Dixon:2020bbt,Dixon:2021tdw,Dixon:2022rse},
\begin{equation}
u = u_1 = \frac{s_{23}}{q^2}\ , \qquad v = u_2 = \frac{s_{13}}{q^2}\ , \qquad w = u_3 = \frac{s_{12}}{q^2}\ ,
\end{equation}
with $q^2 = (p_1+p_2+p_3)^2 = s_{123} = s_{12} + s_{23} + s_{13}$, and so $u+v+w = 1$.

\subsection{BDS-like normalization and space of functions $\cC$}
\label{sec:c_space}

In this subsection we describe our framework for constructing spaces of multiple
polylogarithms (MPLs) which will contain the $\trphi3$ form factor.  We also discuss a specific normalization for the $\trphi3$ form factor which allows us to place it into a particularly small MPL space, called $\cC$, in which the $\trphi2$ form factor also resides.

As was found previously for six-point scattering
amplitudes~\cite{Caron-Huot:2016owq,Caron-Huot:2019vjl,Caron-Huot:2019bsq},
seven-point amplitudes~\cite{Dixon:2016nkn},
and the three-point form factor of
$\trphi2$~\cite{Dixon:2020bbt,Dixon:2022rse}
in planar ${\cal N}=4$ SYM,
it proves useful to work with, not the remainder function $R_{3,3}$,
but a more minimally-infrared subtracted object which lives
in a smaller function space.  Six- and seven-point amplitudes obey Steinmann
relations~\cite{Steinmann,Steinmann2} which forbid double discontinuities
in overlapping channels containing three-particle Mandelstam invariants.
There is a unique normalization (up to a constant overall factor) that
preserves the Steinmann relations.  The (extended) Steinmann relations,
in turn, restrict which letters can be adjacent in the symbol,
thereby reducing the size of the MPL function space needed to describe the amplitudes.

There do not appear to be any direct consequences of the Steinmann relations
for two-particle invariants in all-massless theories. The kinematical variables of the three-point form factor of $\trphi2$ only include two-particle invariants (aside from the operator $q^2=s_{123}$). Nevertheless, improved normalizations were found for this form factor~\cite{Dixon:2020bbt,Dixon:2022rse}. In particular, the normalization adopted in ref.~\cite{Dixon:2022rse}
led to a particularly small space of MPLs, called $\cC$, due to a set of symbol-letter adjacency restrictions.  We wish to employ the space $\cC$ here too, i.e.~for the $\trphi3$ form factor as well as the $\trphi2$ form factor.

First recall that the symbol~\cite{Goncharov:2010jf}
of a weight $n$ MPL function $F$ can be defined
recursively via its total differential,
\begin{equation}
dF = \sum_{\phi_i \in {\cal L}} F^{\phi_i} d\ln\phi_i\ \quad \Rightarrow \quad
 {\cal S}[F] = \sum_{\phi_i \in {\cal L}} {\cal S}[F^{\phi_i}] \otimes \phi_i \,,
\label{symbol_def}
\end{equation}
where the $\phi_i$ are the {\it letters} in the {\it symbol alphabet}
${\cal L}$, and the weight $n-1$ MPLs $F^{\phi_i}$ are referred to
as the $\{n-1,1\}$ coproducts of $F$.  Iterating \eqn{symbol_def} leads
to the rank $n$ tensor,
\begin{equation}
  {\cal S}[F] = \sum_{i_1,i_2,\ldots,i_n} F^{\phi_{i_1},\phi_{i_2},\ldots,\phi_{i_n}}\
           \phi_{i_1} \otimes \phi_{i_2} \otimes \cdots \otimes \phi_{i_n} \,,
\label{S_F}
\end{equation}
where the tensor coefficients $F^{\phi_{i_1},\phi_{i_2},\ldots,\phi_{i_n}}$ are
rational numbers.  We expect polylogarithmic planar ${\cal N}=4$ SYM
amplitudes and form factors to have weight $2L$ at $L$ loops.

The Feynman integrals for the perturbative evaluation of the
$\trphi3$ form factor always have a topology which is
simpler than the most complex integrals for the $\trphi2$ form factor.
The reason is that three massless legs must emerge from the $\trphi3$
vertex, versus two massless legs for $\trphi2$.
Therefore the $\trphi3$ graphs have one fewer propagator denominator
than the maximum possible for $\trphi2$.
As a result, we expect the MPL function space for $\trphi3$
to lie in the same space for $\trphi2$.
That space has six symbol letters,
\be
   {\cal L}_u = \{u,v,w,1-u,1-v,1-w\} \,.
\label{alphabet_uvw}
\ee
However, it is better to switch to a different, multiplicatively
independent set of six letters~\cite{Dixon:2022rse}, for which the adjacency
restrictions for the $\trphi2$ form factor are more apparent:
\be
   {\cal L}_a = \{a,b,c,d,e,f\} \,,
\label{alphabet_abc}
\ee
where
\bea
a &=& \sqrt{\frac{u}{vw}} \,, \qquad b = \sqrt{\frac{v}{wu}} \,,
\qquad c = \sqrt{\frac{w}{uv}} \,, \nonumber\\
d &=& \frac{1-u}{u} \,, \qquad e = \frac{1-v}{v} \,,
\qquad f = \frac{1-w}{w} \,.
\label{abcuvw}
\eea
In fact, ref.~\cite{Dixon:2022rse} used
$a = u/(vw)$, $b = v/(wu)$, $c=w/(uv)$.  The reason for switching
to the normalization in \eqn{abcuvw} is to make the symbol entries
for the $\trphi2$ and $\trphi3$ form factors not just rational numbers, but \emph{integers}.  (See ref.~\cite{Cai:2024znx} and section~\ref{sec:sewing}.)

There is a dihedral $D_3 \equiv S_3$ symmetry of the form factor,
which is generated by
\begin{align}
\text{cycle: } &\{a, b, c, d, e, f\} \to \{b, c, a, e, f, d\} \,,
\nonumber\\
\text{flip: } &\{a, b, c, d, e, f\} \to \{b, a, c, e, d, f\} \,.
\label{cycleflip}
\end{align} 
There are 9 integrability relations for the alphabet ${\cal L}_a$,
which are restrictions on adjacent pairs of letters in either the symbol
or the (function level) coproduct description of the function space.
In addition, as part of the definition of the space $\cC$
defined in ref.~\cite{Dixon:2022rse}, we impose 6 adjacency restrictions:
the following 6 pairs should never appear next to each other in the symbol,
\be
\{a,d\},\ \{b,e\},\ \{c,f\},\ \{d,e\},\ \{e,f\},\ \{f,d\}.
\label{nonadj}
\ee
Or equivalently, in terms of double coproducts of generic functions $F$ in the space,
\be
0 = F^{a,d} = F^{d,a} = F^{d,e} = F^{e,d} \,,
\label{nonadj_alt}
\ee
and cyclic images of these restrictions.  The 6 antisymmetric parts
of these equations, e.g.~$F^{a,d}-F^{d,a}=0$, are actually 6 of the 9
integrability relations.  The other three are
\begin{align}
&F^{a,b}+F^{a,c}-F^{b,a}-F^{c,a} = 0\ , \label{extrapair1}\\
&F^{c,a}+F^{c,b}-F^{a,c}-F^{b,c} = 0\ , \label{extrapair2}\\
&F^{d,b}-F^{d,c}-F^{b,d}+F^{c,d}+F^{e,c}-F^{e,a}-F^{c,e}+F^{a,e}
+F^{f,a}-F^{f,b}-F^{a,f}+F^{b,f} \nonumber\\
&\hskip0.5cm\null + 2\,( F^{c,b}-F^{b,c} ) = 0\ . \label{extrapair3} 
\end{align}
Given these $9\,+\,6$ constraints on the $6 \times 6 = 36$ pairs
of adjacent letters, there are only 21 adjacent pairs of letters in $\cC$.

There are also four independent constraints on adjacent triplets in $\cC$,
of the form~\cite{Dixon:2022rse}
\bea
F^{a,a,b} + F^{a,b,b} + F^{a,c,b} = 0\, ,
\label{nonadjtriple}
\eea
and dihedral images\footnote{Naively, there are six such relations, but only
four remain independent after taking into account that the pair relations also generate triplet relations when tensored with any letter on either side of the pair.}.

For the $\trphi2$ form factor, both the pair adjacency and the triplet relations~(\ref{nonadjtriple}) follow from antipodal duality and the extended Steinmann properties of the hexagon function space describing the six-point (MHV) amplitude.  We currently have no understanding of why the pair and triplet adjacency relations should hold for the $\trphi3$ form factor, but because it lives inside the space $\cC$ it is tempting to conjecture that it should also participate in some kind of antipodal duality.

A BDS-like normalized $\trphi3$ form factor $\EE_{3,3}$ can be defined by
\be
{\cal F}_{3,3} = {\cal F}_{3,3}^{\rm BDS-like} \times \EE_{3,3} \,,
\label{BDSlike1}
\ee
where
\be
{\cal F}_{3,3}^{\rm BDS-like} = {\cal F}_{3,3}^{\rm MHV,\,tree} \times
\exp\biggl\{ \sum_{L=1}^\infty g^{2L}
\biggl[ \Bigl( \frac{\Gcusp^{(L)}}{4}\ +\ {\cal O}(\e) \Bigr)
              M(L\e)\ +\ C_{3,3}^{(L)} \biggr] \biggl\}\ ,
\label{BDSlike2}
\ee
with
\be
M(\e) \equiv M_{3,3}(\e) - \EE_{3,3}^{(1)} \,.
\label{Mdef}
\ee
The loop expansions of $\EE_{3,3}$ and the remainder function $R_{3,3}$ are given by
\bea
\EE_{3,3}(g) &=& 1 + \sum_{L=1}^\infty g^{2L} \, \EE_{3,3}^{(L)} \,,
\label{EE33expandg}\\
R_{3,3}(g) &=& \sum_{L=2}^\infty g^{2L} \, R_{3,3}^{(L)} \,.
\label{R33expandg}
\eea
Notice that the one-loop ($g^2$) term in the expansion of ${\cal F}_{3,3}$ in eq.~(\ref{BDSlike1}) reproduces $M_{3,3}(\e) + C_{3,3}^{(1)}$, which matches eq.~(\ref{RemainderFnDef}) since $R_{3,3}^{(1)} = 0$.

Comparing eqs.~(\ref{RemainderFnDef}) and (\ref{BDSlike1}) at arbitrary orders in $g$ then leads to a relation between the BDS-like normalized form factor and the remainder function,
\be
\EE_{3,3}
= \exp\biggl[ \frac{\Gcusp}{4}\,\EE_{3,3}^{(1)} + R_{3,3} \biggr] \,.
\label{EEtoR}
\ee
The remainder function $R_{3,3}$ has a well-defined relationship to the Wilson loop entering the FFOPE, which we use to fix our boundary conditions. However, we can adjust the finite part of the one-loop $\trphi3$ three-point form factor, $\EE_{3,3}^{(1)}$, and through eq.~(\ref{EEtoR}) this adjustment will change all the higher-loop functions $\EE_{3,3}^{(L)}$ for $L>1$.

Next we adjust $\EE_{3,3}^{(1)}$ in order to make the higher-loop BDS-like normalized form factor $\EE_{3,3}$ satisfy the pair and triplet relations~(\ref{nonadj_alt})
and (\ref{nonadjtriple}) that are required for it to lie in $\cC$.
The symbol of the two-loop form factor~\cite{Brandhuber:2014ica}
begins and ends with the letters $a = a_1$, $b = a_2$, and $c = a_3$.
If we want to preserve this property along with dihedral invariance, then
$\EE_{3,3}^{(1)}$ can only be a linear combination of three quantities,
\be
\EE_{3,3}^{(1)} = c_1 \sum_{i=1}^3 \ln^2 a_i\ +\ c_2 \sum_{i=1}^3 \ln a_i \ln a_{i+1}  \ + \ c_3 \, \zeta_2 \,.
\label{EE3_1_ansatz}
\ee
We might also have added $\sum_{i=1}^3 {\rm Li}_2(1-1/u_i)$ because it lies in $\cC$ and is dihedrally invariant; however, its symbol ends with the letters $d, e, f$, rather than $a, b, c$, and so adding  it would ruin an observed final-entry condition at lower loops, \eqn{FErelationdef} below.

We find a unique result for $c_1$, $c_2$ and $c_3$ that ensures that $\EE_{3,3}$ in \eqn{EEtoR} belongs to $\cC$:
\bea
\EE_{3,3}^{(1)}(u_i) &=&
\sum_{i=1}^3 \Bigl( - \ln^2 u_i + \ln u_i \, \ln u_{i+1} \Bigr) - 3 \, \zeta_2
\nonumber\\
&=&
\frac{1}{4} \sum_{i=1}^3 \Bigl( - \ln^2 a_i + \ln a_i \, \ln a_{i+1} \Bigr)
- 3 \, \zeta_2 \,.
\label{EE33_1loop}
\eea
We actually need to go to three loops in order
to fully fix $\EE_{3,3}^{(1)}$,
because the symbol of $\bigl[\EE_{3,3}^{(1)}\bigr]^2$ obeys both
eqs.~(\ref{nonadj_alt}) and (\ref{nonadjtriple}).

A few other constraints were imposed on $\cC$ in ref.~\cite{Dixon:2022rse}, namely:
\begin{enumerate}
\item $\zeta_2$ was not an independent element of $\cC$; instead, the coefficient of $\zeta_2$ was locked to the other weight two
functions in the space.
\item Three functions were removed from $\cC$ at weight 4 because they did not appear in the coproducts of the $\trphi2$ form factor at high loop orders.
\end{enumerate}
We find that these constraints are obeyed by the $\trphi3$
form factor, at least through six loops. The second constraint fixes
the coefficient of $\zeta_2$ in \eqn{EE33_1loop} to be the value shown.

\renewcommand{\arraystretch}{1.25}
\begin{table}[!t]
\centering
\begin{tabular}[t]{l c c c c c c c c c c c c c c}
\hline\hline
weight $n$
& 0 & 1 &  2 &  3 &  4 &   5 &   6 &    7 &    8 &   9  &  10 & 11
\\\hline\hline
symbols in ${\cal H}$
& 1 & 3 &  6 & 13 & 26 &  51 &  98 &  184 &  339 & 612 & 1083 & 1885 \\\hline
symbols in $\cM$
& 1 & 3 &  9 & 27 & 81 & 243 & 729 & 2187 & 6561 & 19683 & 59049 & 177147
\\\hline
symbols in $\cC$
& 1 & 3 &  9 & 21 & 48 & 108 & 246 &  555 & 1251 &  ?? &  ?? &  ?? \\\hline
functions in $\cC$
& 1 & 3 &  9 & 22 & 52 & 122 & 284 &  654 & 1495 &  ?? &  ?? &  ?? \\\hline
\hline
\end{tabular}
\caption{As a function of the weight, we compare the number of
  \emph{symbols} in the form factor space $\cC$ to the number of symbols in the current hexagon function space ${\cal H}$, and in an earlier form factor space $\cM$. We also
  give the number of {\it functions} in $\cC$, which is larger because
  it includes lower-weight functions multiplied by zeta-values.}
\label{tab:CvsHsymbol}
\end{table}

The construction of $\cC$ was described in some detail in
ref.~\cite{Dixon:2022rse}, and so we do not repeat it here.
In table~\ref{tab:CvsHsymbol}, we compare the dimension of $\cC$
with the dimension of two other polylogarithmic spaces, as a function
of the weight, at symbol level. We denote by $\cC^{(n)}$ the weight-$n$ part of $\cC$, and so on. The hexagon function space $\cH$ was used to bootstrap six-gluon amplitudes to seven loops~\cite{Caron-Huot:2019bsq,Caron-Huot:2019vjl}.
The space $\cM$ contains $\cC$, and it was used to bootstrap the $\trphi2$ form factor to five loops~\cite{Dixon:2020bbt}. It includes the $F^{d,e}=0$ restriction (plus dihedral images of this relation) but {\it not} the $F^{a,d} = 0$ restriction. The number of symbols in $\cM^{(n)}$ is simply $3^n$.

Generic two-loop four-point Feynman integrals with massless internal lines and only one external mass~\cite{Gehrmann:2000zt,Gehrmann:2001ck} lie within $\cM$~\cite{Dixon:2020bbt}.  Hence $\cM$ also contains the transcendental functions for generic two-loop amplitudes in quantum chromodynamics (QCD) with massless quarks and one external massive ``operator" (for example, a Higgs boson coupling to gluons through an effective operator, or electroweak vector bosons coupling to quarks).
At three loops, some integrals with the relevant topologies do {\it not} lie in $\cM$~\cite{Henn:2023vbd,Syrrakos:2023mor} (some even have additional letters).
On the other hand, there is a nontrivial cancellation between such integrals in the planar three-loop three-point form factor for the electromagnetic current $\gamma^* \to q g \bar{q}$ in QCD~\cite{Gehrmann:2023jyv}. This observation suggests that the space $\cM$ may still be relevant for QCD beyond two loops.

In any event, table~\ref{tab:CvsHsymbol} shows that $\cM$ is considerably
larger than $\cC$ at high weights, but that $\cC$ still grows more quickly
than the hexagon function space ${\cal H}$.  We give both the number of
symbols and the number of functions in $\cC$.  The additional ``beyond-the-symbol'' functions
have the form of a zeta value, such as $\zeta_k$ for $k>2$,
multiplied by the function space at $k$ lower weights.
For example, at weight 4 we have to add four such functions,
$\zeta_3 \, \ln a_i$, $i=1,2,3$, and $\zeta_4$.   (As mentioned above,
$\zeta_2$ is treated differently; it is not considered to be
independent of the other weight 2 functions.)


\section{Imposing the constraints}
\label{sec:Constraints}

We expect to find the form factor $\EE_{3,3}^{(L)}$ in the space $\cC$,
so we write an ansatz,
\be
  \EE_{3,3}^{(L)} = \sum_{i=1}^{{\rm dim} \, \cC^{(2L)}} c_i F_i^{(2L)} \,,
\label{ansatz}
\ee
where the $c_i$ are rational numbers and $F_i^{(2L)}$ are basis elements for the
weight $2L$ component $\cC^{(2L)}$ of the space $\cC$.
Once we construct $\cC^{(2L)}$,
the next step is to impose constraints on the function space
until a unique solution is found.  Table~\ref{tab:Ctripparameters}
shows the number of parameters left in our ansatz
after imposing various constraints sequentially.
The first line after the loop order is just the total
number of functions in $\cC$ at weight $2L$, which can also be found
on the corresponding line in table~\ref{tab:CvsHsymbol}.

\renewcommand{\arraystretch}{1.25}
\begin{table}[!t]
\centering
\begin{tabular}[t]{l c c c c c}
\hline\hline
$L$                           &  2 &   3 &   4  &   5  &  6
\\\hline\hline
functions in $\cC$            & 52 & 284 & 1495 & $\sim$8000 & ?????
\\\hline
dihedral symmetry             & 13 &  63 &  302 & $\sim$1400 & ????
\\\hline
$(L-1)$ final entries         &  4 &  15 &   47 &  190 &  407 \\\hline
$(L+1)^{\rm st}$ discontinuity &  3 &   13 &   43 &  182 &  394 \\\hline
OPE $T^1\,\ln^{L}T$           &   2 &  10 &   38 &  171 &  ??? \\\hline
OPE $T^1\,\ln^{L-1}T$         &   1 &   6 &   31 &  158 &  ??? \\\hline
OPE $T^1\,\ln^{L-2}T$         &   0 &   2 &   20 &  137 &  322 \\\hline
OPE $T^1\,\ln^{L-3}T$         &   0 &   0 &    4 &  103 &  272 \\\hline
OPE $T^1\,\ln^{L-4}T$         &   0 &   0 &    0 &   50 &  190 \\\hline
OPE $T^1\,\ln^{L-5}T$         &   0 &   0 &    0 &   0 &   64 \\\hline
OPE $T^1\,\ln^{L-6}T$         &   0 &   0 &    0 &   0 &   0 \\\hline\hline
\end{tabular}
\caption{Number of parameters left when bootstrapping the form factor
  $\EE_{3,3}^{(L)}$ in the function space $\cC$
  at full function level. We use the conditions on the final $(L-1)$ entries,
  because they can be deduced at $(L-1)$ loops, as well as the discontinuity
  and OPE information shown.}
\label{tab:Ctripparameters}
\end{table}


The form factor should be dihedrally invariant.  We have classified how all of the functions in $\cC$ transform under the cycle and flip generators of $D_3$ given in \eqn{cycleflip}. It is then easy to find the dihedrally invariant subspace. According to the relevant line in table~\ref{tab:Ctripparameters}, this subspace has a dimension that is a bit larger than 
$1/|D_3| = 1/6$ of the dimension of the full space.

The second set of constraints we impose are {\it multi-final-entry conditions}. These conditions are empirical, and are deduced from the lower-loop form-factor results, once the space of their
multiple coproducts has stabilized, i.e.~the dimension does not change with additional loop orders.
For example, by examining the two-loop remainder function~\cite{Brandhuber:2014ica}, and the one-loop form factor given in \eqn{EE33_1loop}, we find that they both obey the four final-entry relations,
\bea
\EE_{3,3}^{d} \,=\, \EE_{3,3}^{e} \,=\, \EE_{3,3}^{f} &=& 0 \,,
\label{FErelationdef}\\
\EE_{3,3}^{a}\ +\EE_{3,3}^{b}\ +\EE_{3,3}^{c} &=& 0 \,.
\label{FErelationabc}
\eea
Here $\EE_{3,3}^{\phi}$ refers to a $\{2L-1,1\}$ coproduct of the $L$-loop, weight-$2L$ function $\EE_{3,3}^{(L)}$ (see eq.~(\ref{symbol_def})). 
We assume that \eqns{FErelationdef}{FErelationabc} hold to all loop orders. These linear relations reduce the number of weight $2L-1$
single final entries $\EE_{3,3}^{\phi}$ from six to two,
which we can take to be $\EE_{3,3}^{a}$ and $\EE_{3,3}^{b}$.

Then we take the six coproducts of $\EE_{3,3}^{a}$ and $\EE_{3,3}^{b}$, i.e.~$\EE_{3,3}^{\phi,a}$ and $\EE_{3,3}^{\phi,b}$, where $\phi\in\cL_a$, and we count how many of the $6 \times 2 = 12$ such weight $2L-2$ double final entries are linearly independent.  By three loops, this dimension has stabilized at six, which lets us identify six double-final-entry relations,
\bea
\EE_{3,3}^{d,a} &=& \EE_{3,3}^{e,b} = 0, \qquad
\EE_{3,3}^{f,b}  = -\,\EE_{3,3}^{f,a} \,, \qquad
\EE_{3,3}^{c,a}  = -\,\EE_{3,3}^{a,a} - \EE_{3,3}^{b,a} \,, \nonumber\\
\EE_{3,3}^{c,b} &=& -\,\EE_{3,3}^{a,b} - \EE_{3,3}^{b,b} \,, \qquad
\EE_{3,3}^{d,b}  = \EE_{3,3}^{e,a} - \EE_{3,3}^{f,a}
                + \EE_{3,3}^{a,b} - \EE_{3,3}^{b,a} \,.
\label{doubleFErelations}
\eea
We incorporate these relations, and the corresponding final triple, quadruple, etc., relations, into our ansatz for the form factor as soon as they stabilize or {\it saturate}, i.e.~once the appropriate spaces of coproducts have the same dimension at two consecutive loop orders. (See table~\ref{tab:EEcopdimsymb} below.)

The multi-final-entry relations are extremely useful computationally,
because we do not have to construct the space $\cC$ all the way up to
weight $2L$; we can ``sew'' a lower-weight component of $\cC$ to
a representation of the allowed final entries.
In practice, we have constructed the space $\cC$ to weight 8, which is sufficient
to directly represent the four-loop form factor.  At five loops, we
write the weight 10 form factor in terms of its weight 8 double coproducts,
or $\{8,1,1\}$ coproducts.  At six loops, we use the $\{8,1,1,1,1\}$
(quadruple) coproducts.  Knowing all the (expected) relations between
the multiple final entries directly cuts the size of the initial ansatz
and thus the computational work required.  We will discuss the sewing matrix representation for the symbol of the final result in sect.~\ref{sec:sewing}.

The third set of constraints comes from the general structure of the FFOPE
described in section~\ref{sec:FFOPE}.  \Eqn{FFOPE_def} restricts
the logarithmic discontinuities of the form factor Wilson loop $\mathcal{W}_{3,3}$~\cite{Gaiotto:2011dt}. We analyze the leading dependence on $\tau = - \ln T$, as $\tau\to\infty$.  This dependence constrains the discontinuities in $w$ through eq.~(\ref{OPEparam3}), since $\ln w \sim -2\ln T \sim -2\tau$ as $\tau\to\infty$ and $w\to0$.
From \eqn{FFOPE_def}, the maximum power of $\tau$ available at $L$ loops is $L$, which comes from series expanding the one-loop energy $g^2 \, E_\psi^{(1)}$ down out of the exponential to get a factor of $g^{2L} \, \tau^L$. Therefore
\be
[{\rm Disc}_w ]^{L+1} \mathcal{W}_{3,3}^{(L)} = 0\,.
\label{DiscW}
\ee
This does not mean that the $(L+1)^{\rm st}$ discontinuity of
$\EE_{3,3}^{(L)}$ is zero, however, because \eqn{W33} features
a relative factor of $\exp[4g^2\tau^2]$ between 
$\mathcal{W}_{3,3}$ and $\EE_{3,3}$, which has {\it two}
discontinuities per loop.  Nevertheless, the $(L+1)^{\rm st}$ discontinuity
of $\EE_{3,3}^{(L)}$ is computable in terms of lower-loop multiple
discontinuities, by taking the logarithm of \eqn{W33} and using \eqn{DiscW}
to show that
\be
[{\rm Disc}_w ]^{L+1} [\ln \mathcal{W}_{3,3}]^{(L)} = 0\,.
\label{DisclnW}
\ee
Since $[\ln \mathcal{W}_{3,3}]^{(L)}$ and $[\ln \EE_{3,3}]^{(L)}$ differ by something proportional to $\tau^2$, we find that
\be
[{\rm Disc}_w ]^{L+1} [\ln \EE_{3,3}]^{(L)} = 0   \qquad (L>1)\,.
\label{DisclnEE}
\ee
Next we insert the perturbative expansion~(\ref{EE33expandg}) of $\EE_{3,3}$ into \eqn{DisclnEE}, and use the Leibniz rule to apply discontinuities to products of lower-loop expressions.  In this way, we rewrite \eqn{DisclnEE} in terms of the desired quantity, $[{\rm Disc}_w ]^{L+1} \EE_{3,3}^{(L)}$, and lower-loop contributions. The expression is lengthy and unenlightening.
Once $[{\rm Disc}_w ]^{L+1} \EE_{3,3}^{(L)}$ is computed in this way, matching it
to the multiple discontinuity of the ansatz leaves the numbers of parameters shown in the fourth line
of table~\ref{tab:Ctripparameters}.

The fourth step is to expand around the near-collinear limit and
utilize the FFOPE data for the leading OPE correction, whose construction is described in
section~\ref{sec:FFOPE}.
These data are provided as the $T^1$ terms in the $T = e^{-\tau} \to 0$ limit, accompanied by logarithms of $T$ (powers of $\tau$) up to $\ln^L T$ at $L$ loops. Also, the dependence on $S=e^\sigma$ is given as a series expansion in $S$ as $S\to0$, accompanied by logarithms of $S$. So we have to expand the ansatz first in $T$, and then in $S$, in order to match the FFOPE data.  

As seen in table~\ref{tab:Ctripparameters}, it takes quite a bit of data to fix all the parameters.  Note that the single power of $T$ listed there comes from a tree-level rational prefactor, $\Omega_3$ in \eqn{eq:Omega3-data}.  Thus, to match the leading OPE terms, the transcendental function $\EE_{3,3}^{(L)}$ only has to be expanded to leading power in the collinear limit\footnote{In contrast, for the leading OPE terms in the $\trphi2$ form factor, the transcendental function $\EE^{(L)}$ must be expanded to a subleading power in $T$.}. Nevertheless, one can see that all logarithms at this power are
required; that is, one only gets to 0 parameters left by using
the final $T^1 \ln^0 T$ layer of leading OPE information.  A couple of 6 loop entries are missing in table~\ref{tab:Ctripparameters} because at 6 loops the parameters were fixed in a different order, i.e.~all symbol-level parameters for all subleading logs were fixed first, followed by the function-level parameters.

We also expanded the fully-determined results to order $T^3$, and checked them against the subleading OPE data described in section~\ref{sec:FFOPE} and appendix~\ref{appendix_1}.
From the fully-determined results, we can provide the ``resummed'' version of the OPE data, where the functions of $S$ are given exactly and not as series expanded in $S$. These results are described in section~\ref{sec:resumOPE}.

Because the symbol letters are linear in $u$ and $v$, the $L$-loop $\trphi3$ form factor $\EE_{3,3}^{(L)}(u,v,w)$ can be expressed for arbitrary $(u,v,w)$ in terms of $G$ functions with argument $u$ and indices $a_i \in \{0,1,v,1-v\}$, and $G$ functions with argument $v$ and indices $a_i \in \{0,1\}$.  We provide such a representation through $L=5$ in the ancillary file {\tt EE33G.txt}.


\section{Properties of the results}
\label{sec:Properties}

Having determined the $\trphi3$ form factor $\EE_{3,3}^{(L)}$
through 6 loops, we now proceed to examine some of its properties,
starting with what it tells us about the multi-final-entry conditions.

\subsection{Structural results}
\label{sec:structural}

Once we determine $\EE_{3,3}^{(L)}$, we can differentiate it and
determine how many of its multi-derivatives are independent.
More precisely, we determine the dimensions of its spaces of
independent $\{n,1,\ldots,1\}$ coproducts, in order to see
how they saturate with the loop order.  The $\trphi2$ form factor
was analyzed similarly through eight loops~\cite{Dixon:2022rse}.
We can perform this analysis both at symbol level and at function level.

\renewcommand{\arraystretch}{1.25}
\begin{table}[!t]
\centering
\begin{tabular}[t]{l c c c c c c c c c c c c c}
\hline\hline
weight $n$
& 0 & 1 & 2 & 3 & 4 &  5 &  6 &  7 &  8 &  9 & 10 & 11 & 12\\\hline\hline
$L=1$
& \green{1} & \blue{2} & \blue{1} &  &  &  &  &  &  &  & & &
\\\hline
$L=2$
& \green{1} & \green{3} & 3 & \blue{2} & \blue{1} & & & & & & & &
\\\hline
$L=3$
& \green{1} & \green{3} & \green{9} & \blue{13} & \blue{6} & \blue{2} & \blue{1}
&  &  &  & & &
\\\hline
$L=4$
& \green{1} & \green{3} & \green{9} & \green{21} & \blue{29} & \blue{13}
& \blue{6} & \blue{2} & \blue{1} &  & & &
\\\hline
$L=5$
& \green{1} & \green{3} & \green{9} & \green{21} &  \green{48} & \blue{57}
& \blue{29} & \blue{13} & \blue{6} & \blue{2} & \blue{1} & &
\\\hline
$L=6$
& \green{1} & \green{3} & \green{9} & \green{21} & \green{48} & 105
& \red{112} & \blue{57} & \blue{29} & \blue{13} & \blue{6} & \blue{2}
& \blue{1}
\\\hline\hline
\end{tabular}
\caption{The number of independent $\{n,1,1,\ldots,1\}$ coproducts
  of the form factor $\EE_{3,3}^{(L)}$ through $L=6$ loops, at
  \emph{symbol} level.  The meaning of the colors is discussed in the text.}
\label{tab:EEcopdimsymb}
\end{table}

The symbol-level dimensions are shown in table~\ref{tab:EEcopdimsymb}.
A \green{green} number in the table denotes saturation of the space
$\cC$ constructed from the bottom up. This saturation is indicated
when the $(L,n)$ number is the same as the $(L+1,n)$ number right below it.
If the number saturates below the number of functions that we constructed
directly, then we can identify and remove extra functions which are not
needed at a given weight. Such removal of functions also impacts higher weights as well,
since their coproducts are restricted.  A \blue{blue} number denotes
saturation of the space of multi-final-entries.  It is indicated by the
$(L,n)$ number being the same as the $(L+1,n+2)$ number.
Once such a number saturates, we identify the multi-final-entry relations
that are responsible for that number of independent entries.
A red number means that the number is presumably saturated,
but the evidence of two successive loop orders is not quite there yet.
For example, the number of independent hextuple coproducts is strongly
expected to be \red{112}, given the pattern at lower loops of when
this number saturates, namely at $n=L$.

Also, in table~\ref{tab:EEcopdimsymb}
there are only \green{48} independent symbols at weight 4,
from the 5 and 6 loop coproducts at symbol level.
The weight 5 symbols have not yet quite saturated at 6 loops,
relative to what we know from the 8 loop $\trphi2$ form factor,
which is that there should be \green{108} weight 5 symbols~\cite{Dixon:2022rse}.

Table~\ref{tab:EEcopdim} gives the same numbers at function level.
The numbers on the right-side are the same as in table~\ref{tab:EEcopdimsymb}.
In other words, the multi-final-entry analysis can be done at
either symbol level or function level.
On the left-hand side, the numbers are a bit larger, accounting
for beyond-the-symbol functions with zeta-value prefactors in $\cC$.
Note that the weight 4 number \green{51} found at 5 and 6 loops
is one smaller than the weight 4 dimension of $\cC$.
This was also the weight 4 number encountered at 7 and 8 loops
for $\trphi2$~\cite{Dixon:2022rse}, and the same function is ``missing''
in all cases:  $\zeta_3 \ln(abc)$.

\renewcommand{\arraystretch}{1.25}
\begin{table}[!t]
\centering
\begin{tabular}[t]{l c c c c c c c c c c c c c}
\hline\hline
weight $n$
& 0 & 1 & 2 & 3 & 4 &  5 &  6 &  7 &  8 &  9 & 10 & 11 & 12\\\hline\hline
$L=1$
& \green{1} & \blue{2} & \blue{1} &  &  &  &  &  &  &  & & &
\\\hline
$L=2$
& \green{1} & \green{3} & 3 & \blue{2} & \blue{1} & & & & & & & &
\\\hline
$L=3$
& \green{1} & \green{3} & \green{9} & \blue{13} & \blue{6} & \blue{2} & \blue{1}
&  &  &  & & &
\\\hline
$L=4$
& \green{1} & \green{3} & \green{9} & \green{22} & \blue{29} & \blue{13}
& \blue{6} & \blue{2} & \blue{1} &  & & &
\\\hline
$L=5$
& \green{1} & \green{3} & \green{9} & \green{22} & \green{51} & \blue{57}
& \blue{29} & \blue{13} & \blue{6} & \blue{2} & \blue{1} & &
\\\hline
$L=6$
& \green{1} & \green{3} & \green{9} & \green{22} & \green{51} & 112
& \red{112} & \blue{57} & \blue{29} & \blue{13} & \blue{6} & \blue{2}
& \blue{1}
\\\hline\hline
\end{tabular}
\caption{The number of independent $\{n,1,1,\ldots,1\}$ coproducts
  of the form factor $\EE_{3,3}^{(L)}$ through
  $L=6$ loops, at \emph{function} level.  The colors have the same
  meaning as in the symbol-level table~\ref{tab:EEcopdimsymb}.}
\label{tab:EEcopdim}
\end{table}

In table~\ref{tab:EEfinalentryconditions} we record, in the middle row,
the number of independent final $w$ entries, extracted from
tables~\ref{tab:EEcopdimsymb} and \ref{tab:EEcopdim}.  The upper row is
the number of final $w$ entries that would have followed from the
$(w-1)$ final-entry conditions, and the generic pair
relations~(\ref{nonadj_alt})--(\ref{extrapair3})
and generic triplet relations~(\ref{nonadjtriple}).
The number in the bottom row is the difference, i.e.~the number of
new weight $w$ final-entry conditions.  One can see that it jumps significantly starting at $w=5$.
The single and double final entry relations are given in
eqs.~(\ref{FErelationdef})--(\ref{doubleFErelations}).
In an ancillary file {\tt phi3FFmultifinalentry.txt} we provide the
multi-final-entry relations for $\trphi3$ for $w$ up to 6.

\renewcommand{\arraystretch}{1.25}
\begin{table}[!t]
\centering
\begin{tabular}[t]{l c c c c c c}
\hline\hline
$w$                                     &  1  &  2 &  3  &  4  &  5  &  6
\\\hline\hline
before final $w$ entry conditions       &  6  &  7 &  15 &  30 &  66 & 120
\\\hline
after final $w$ entry conditions        &  2  &  6 &  13 &  29 &  57 & 112
\\\hline
new final $w$ entry conditions          &  4  &  1 &   2  &  1  &  9 &   8
\\\hline
\end{tabular}
\caption{Dimension of the space of potential $\{2L-w,1,\dots,1\}$ coproduct entries of the $L$-loop form factor for $\trphi3$, before and after imposing the $w$-final-entry conditions. We also give their difference, which is the number of non-trivial, new constraints at weight $w$ (not accounting for dihedral symmetry).}
\label{tab:EEfinalentryconditions}
\end{table}

The numbers of terms in the symbol of the $\trphi3$ form factor $\EE_{3,3}^{(L)}$ are shown in table~\ref{tab:symbolterms} through six loops, along with the corresponding numbers for the $\trphi2$ form factor $\EE^{(L)}$. We observe that the symbols are actually considerably lengthier for $\trphi3$ than they are for $\trphi2$, even though the integrals are simpler in principle, as mentioned in section~\ref{sec:c_space}. Of course, the symbol length depends on the alphabet used, but it does not seem likely that there is a better alphabet than ${\cal L}_a$, given all the adjacency restrictions it features.

\renewcommand{\arraystretch}{1.25}
\begin{table}[!t]
\centering
\begin{tabular}[t]{l c c c c c c c c c c}
\hline\hline
loop order $L$
& 1 &  2 &   3 &      4 &       5 &         6  \\\hline
terms in ${\cal S}[\EE^{(L)}]$
& 6 & 12 & 636 & 11,208 & 263,880 & 4,916,466 \\\hline
terms in ${\cal S}[\EE_{3,3}^{(L)}]$
& 9 & 105 & 1773 & 44,391 & 747,837 & 14,637,501 \\\hline\hline
\end{tabular}
\caption{Comparison between the number of terms in the symbols of the $\trphi2$ and $\trphi3$ form factors,
$\EE^{(L)}$ and $\EE_{3,3}^{(L)}$ respectively, 
as a function of the loop order $L$ through six loops.}
\label{tab:symbolterms}
\end{table}

\subsection{Sewing matrix representation}
\label{sec:sewing}

The symbol provides a large part of the analytic answer.
Once the symbol alphabet is specified, it is canonical and basis independent.
Also, using the alphabet ${\cal L}_a$ in \eqn{alphabet_abc}, all
the coefficients of the symbol of $\EE_{3,3}^{(L)}$
turn out to be not just rational numbers, but actually integers,
through six loops\footnote{The same is also true for the symbol of $\EE^{(L)}$, where all coefficients are actually divisible by 4, starting at two loops, all the way through 8 loops~\cite{Cai:2024znx}.}. We provide the symbol in the ancillary file {\tt phi3FFsymbols.txt} through 6 loops.  Table~\ref{tab:symbolterms} shows that the symbol can be quite lengthy.
However, we know that there are many correlations between its entries, such as the pair and triplet relations, eqs.~(\ref{nonadj_alt})--(\ref{nonadjtriple}).
In this section,
we discuss a way to compress the information in the symbol, while also aiming to maintain the integrality of coefficients.

The general idea is to represent the symbol in terms of a ``sewing matrix''
which connects a basis for the front part of the symbol with a basis
for the back part of the symbol.
The front-space basis is based on the construction of the space $\cC$,
while the back-space basis is based on solving the multi-final entry
conditions.  In particular, when comparing the $\trphi2$ form factor $\EE$ to the $\trphi3$
form factor $\EE_{3,3}$, they will share the same front-space,
but have different back-spaces, because their multi-final-entry relations
are different.  We write,
\bea
{\cal S}[\EE^{(L)}] &=& \sum_{i,j} F_i^{(w)} M_{ij}^{(w,w')} B_j^{(w')} \,,
\label{Sewphi2}\\
{\cal S}[\EE_{3,3}^{(L)}]
&=& \sum_{i,j} F_i^{(w)} \hat{M}_{ij}^{(w,w')} \hat{B}_j^{(w')} \,.
\label{Sewphi3}
\eea
Here $F_i^{(w)}$ is a basis for the front-space at weight $w$, i.e.~it is a basis for $\cC^{(w)}$, the weight-$w$ part of $\cC$;
whereas $B_j^{(w')}$ ($\hat{B}_j^{(w')}$) is a basis for the back-space for the form factor for $\trphi2$ ($\trphi3$).  Finally,
$M_{ij}^{(w,w')}$ ($\hat{M}_{ij}^{(w,w')}$) is the corresponding
\emph{sewing matrix}.
In the simplest case, $w' = 2L-w$, the matrix elements of $M_{ij}^{(w,2L-w)}$ are rational numbers.
They depend on the bases used for the front- and back-spaces.  We would like the symbols for all elements of the front- and back-space bases to have only integer coefficients.
We conjecture that by choosing these integral bases carefully, the matrix elements $M_{ij}^{(w,2L-w)}$ can all be made integers as well, for any $L$ and $w$.

When $w+w'<2L$, the matrix elements are actually symbols with weight $2L-w-w'$.  We refer to this matrix as a {\it generalized sewing matrix}.

For $w' = 2L-w$, the sewing representation is not particularly new; we implicitly use such a representation whenever the weight of the amplitude or form factor exceeds the maximum weight of the function space basis we have constructed. For example, in ref.~\cite{Dixon:2022rse} the maximum weight for $\cC$ that was constructed was 8, but one could sew this space together with back-spaces with weight up to 8, in order to obtain the $\trphi2$ form factor up to 8 loops, or weight 16. Here, we pose a new question: Can one have all $M_{ij}^{(w,2L-w)}\in \mathbb{Z}$, while requiring that the symbol coefficients in $F_i$ and in $B_j$ are also all integers? The answer depends on the basis chosen for $F_i$ and $B_j$, but it seems possible to find such bases, at least up to a certain weight.

Let us now give some low-weight examples.
The weight 1 front space is given by the 3 allowed first entries,
\be
F^{(1)} = \{{\tt a},\ {\tt b},\ {\tt c} \} \,.
\label{frontF1}
\ee
The weight 2 and 3 front spaces have dimensions 9 and 21, respectively,
and their basis elements are
\be
F^{(2)} = \{ {\tt aa},\ {\tt bb},\ {\tt cc},\
{\tt bc+cb},\ {\tt ca+ac},\ {\tt ab+ba},\
{\tt bd+cd},\ {\tt ce+ae},\ {\tt af+bf} \} \,,
\label{frontF2}
\ee
and
\bea
F^{(3)} &=& \{ {\tt aaa},\ {\tt bbb},\ {\tt ccc},\ 
{\tt aab+aba+baa-bba-bab-abb},\ 
\nonumber\\  &&\hskip0cm\null
{\tt bbc+bcb+cbb-ccb-cbc-bcc},\ 
{\tt cca+cac+acc-aac-aca-caa},\
\nonumber\\  &&\hskip0cm\null
{\tt bdd+cdd},\ {\tt cee+aee},\ {\tt aff+bff},\
\nonumber\\  &&\hskip0cm\null
{\tt bdb+cdb+bdc+cdc},\
{\tt cec+aec+cea+aea},\ 
{\tt afa+bfa+afb+bfb},\ 
\nonumber\\  &&\hskip0cm\null
{\tt bbd+bcd+cbd+ccd},\ 
{\tt cce+cae+ace+aae},\ 
{\tt aaf+abf+baf+bbf},\ 
\nonumber\\  &&\hskip0cm\null
{\tt caa+aca-cca-cab-acb+cdc+bdc
  +cec+aec+ccb-bfb-afb-bdb}
\nonumber\\  &&\hskip2cm\null
  {\tt -cdb
  +aab+bfa+cea+aea+afa},\ 
\nonumber\\  &&\hskip0cm\null
{\tt abb+bab-aab-abc-bac+aea+cea
  +afa+bfa+aac-cdc-bdc-cec}
\nonumber\\  &&\hskip2cm\null
 {\tt -aec
  +bbc+cdb+afb+bfb+bdb},\ 
\nonumber\\  &&\hskip0cm\null
{\tt bcc+cbc-bbc-bca-cba+bfb+afb
  +bdb+cdb+bba-aea-cea-afa}
\nonumber\\  &&\hskip2cm\null
  {\tt -bfa
  +cca+aec+bdc+cdc+cec},\ 
\nonumber\\  &&\hskip0cm\null
{\tt bdc+cdc-cec-aec-bbd+aae+ccd-cce},\ 
\nonumber\\  &&\hskip0cm\null
{\tt cea+aea-afa-bfa-cce+bbf+aae-aaf},\ 
\nonumber\\  &&\hskip0cm\null
{\tt afb+bfb-bdb-cdb-aaf+ccd+bbf-bbd}
\} \,.
\label{frontF3}
\eea
We have dropped the tensor products and written the symbols as strings
to save space.  For example, ${\tt bd+cd}$ stands for
$b\otimes d + c\otimes d$.  We provide front-space bases through weight 6 in the ancillary file {\tt Cfrontspace.txt}.   However, only the bases through weight 4 seem to generate integral sewing matrices (although weight 5 is close).

Next, we write the bases for the first few backspaces
for the $\trphi2$ form factor:
\be
B^{(1)} = \{ {\tt d},\ {\tt e},\ {\tt f} \} \,,
\label{backB1}
\ee
\be
B^{(2)} = \{ {\tt dd},\ {\tt ee},\ {\tt ff},\ 
{\tt ae+af},\ {\tt bd+bf},\ {\tt cd+ce} \} \,,
\label{backB2}
\ee
\bea
B^{(3)} &=& \{ {\tt ddd},\ {\tt eee},\ {\tt fff},\
{\tt aae+aaf},\ {\tt bbd+bbf},\ {\tt ccd+cce},\ 
\nonumber\\  &&\hskip0cm\null
{\tt fbd+fbf+dbd+dbf},\
{\tt dce+dcd+ece+ecd},\
{\tt eaf+eae+faf+fae},\
\nonumber\\  &&\hskip0cm\null
{\tt ecd+ece+cee-cdd},\
{\tt fae+faf+aff-aee},\
{\tt dbd+dbf+bdd-bff} \} \,.
\label{backB3}
\eea
These bases solve the multi-final-entry relations given
in ref.~\cite{Dixon:2022rse}.  We provide back-space bases for $\trphi2$ through weight 7 in the ancillary file {\tt phi2backspace.txt}.  

For the $\trphi3$ form factor, the first few backspace bases are given by:
\be
\hat{B}^{(1)} = \{ {\tt a-b},\ {\tt b-c} \} \,,
\label{back3B1}
\ee
\bea
\hat{B}^{(2)} &=& \{ {\tt aa-ab-ba+bb},\
{\tt bb-bc-cb+cc},\
{\tt cc-ca-ac+aa},\
\nonumber\\  &&\hskip0cm\null
{\tt db-dc+ab-ac-cb+cc},\
{\tt ec-ea+bc-ba-ac+aa},\
\nonumber\\  &&\hskip0cm\null
{\tt fa-fb+ca-cb-ba+bb} \} \,,
\label{back3B2}
\eea
and
\bea
\hat{B}^{(3)} &=& \{
{\tt aaa-caa+cca-ccc-aca+acc-aac+cac},\
\nonumber\\  &&\hskip0cm\null
{\tt bbb-abb+aab-aaa-bab+baa-bba+aba},\
\nonumber\\  &&\hskip0cm\null
{\tt ccc-bcc+bbc-bbb-cbc+cbb-ccb+bcb},\
\nonumber\\  &&\hskip0cm\null
{\tt bbb-aac-aca-caa-bba-bab-abb+aaa+cba+bca+ccc-ccb}
\nonumber\\  &&\hskip2cm\null
      {\tt -cbc-bcc+bac+cab+abc+acb},\
\nonumber\\  &&\hskip0cm\null
{\tt bca\!-\!baa\!+\!caa\!+\!bac\!+\!ccc\!-\!bcc\!-\!cca\!-\!cac
\!+\!dba\!-\!dca\!-\!ddb\!-\!dbc\!+\!dcc\!+\!ddc},\
\nonumber\\  &&\hskip0cm\null
{\tt cab\!-\!cbb\!+\!abb\!+\!cba\!+\!aaa\!-\!caa\!-\!aab\!-\!aba
\!+\!ecb\!-\!eab\!-\!eec\!-\!eca\!+\!eaa\!+\!eea},\
\nonumber\\  &&\hskip0cm\null
{\tt abc\!-\!acc\!+\!bcc\!+\!acb\!+\!bbb\!-\!abb\!-\!bbc\!-\!bcb
\!+\!fac\!-\!fbc\!-\!ffa\!-\!fab\!+\!fbb\!+\!ffb},\
\nonumber\\  &&\hskip0cm\null
{\tt bbc-bac+cbb-cbc+cac+bdb-cdb-bdc+cdc-cab-bbb+bab},\
\nonumber\\  &&\hskip0cm\null
{\tt cca-cba+acc-aca+aba+cec-aec-cea+aea-abc-ccc+cbc},\
\nonumber\\  &&\hskip0cm\null
{\tt aab-acb+baa-bab+bcb+afa-bfa-afb+bfb-bca-aaa+aca},\
\nonumber\\  &&\hskip0cm\null
{\tt dcb-dcc+dbc-dbb-abb-bbc-acc-bcb+acb+abc+bbb+bcc},\
\nonumber\\  &&\hskip0cm\null
{\tt eac-eaa+eca-ecc-bcc-cca-baa-cac+bac+bca+ccc+caa},\
\nonumber\\  &&\hskip0cm\null
{\tt fba-fbb+fab-faa-caa-aab-cbb-aba+cba+cab+aaa+abb} \} .~~~~~
\label{back3B3}
\eea
At weights $w'=1,2$, the bases $\hat{B}^{(w')}$ satisfy the single-
and double-final-entry relations,
eqs.~(\ref{FErelationdef})--(\ref{doubleFErelations}).
We provide back-space bases for $\trphi3$ through weight 5 in the ancillary file {\tt phi3backspace.txt}.  

Simple examples of sewing matrices $M^{(w,w')}$ are provided by the symbol of the $\trphi2$ form factor $\EE$, which at one loop in the alphabet~\eqref{alphabet_abc} is
\be
{\cal S} [ \EE^{(1)} ] = -\,2\left( {\tt ae+af+bd+bf+cd+ce} \right)\,.
\label{phi2loop1}
\ee
There are three sewing matrices with $w'=2-w$
that can be constructed from it,
\bea
M^{(0,2)} &=& \begin{bmatrix}
    0 & 0 & 0 & -2 & -2 & -2
\end{bmatrix} \,,
\label{M102} \\
M^{(1,1)} &=& \begin{bmatrix}
 0  & -2 & -2\\
 -2 &  0 & -2\\
 -2 & -2 &  0
\end{bmatrix} \,,
\label{M111} \\
M^{(2,0)} &=& \begin{bmatrix}
    0 & 0 & 0 & 0 & 0 & 0 & -2 & -2 & -2
\end{bmatrix}^T \,,
\label{M120}
\eea
where we wrote the transpose of $M^{(2,0)}$ to save space.  Recall that the loop order of $M^{(w,w')}$ is implicit, as $L = (w+w')/2$. We provide additional examples of sewing matrices for the $\trphi2$ form factor, through 6 loops, in the ancillary file {\tt phi2SewMatrix.txt}.

Now we return to describing the symbols of $\trphi3$ in the sewing matrix formalism.  The one-loop symbol is
\be
{\cal S} [ \EE_{3,3}^{(1)} ] = {\tt ab+ba+bc+cb+ca+ac}-2\left({\tt aa+bb+cc}\right) \,.
\label{phi3loop1}
\ee
The three sewing matrices are
\bea
\hat{M}^{(0,2)} &=& \begin{bmatrix}
    -1 & -1 & -1 & 0 & 0 & 0
\end{bmatrix} \,,
\label{M3102} \\
\hat{M}^{(1,1)} &=& \begin{bmatrix}
 -2 & -1\\
  1 & -1\\
  1 &  2
\end{bmatrix} \,,
\label{M3111} \\
\hat{M}^{(2,0)} &=& \begin{bmatrix}
    -2 & -2 & -2 & 1 & 1 & 1 & 0 & 0 & 0
\end{bmatrix}^T \,.
\label{M3120}
\eea
The two-loop symbol of $\trphi3$ already has 105 terms, so we don't
present it here. (See the ancillary file {\tt phi3FFsymbols.txt}.)
However, the following sewing matrices for it are not too cumbersome: 
\setcounter{MaxMatrixCols}{13}
\bea\label{M3213}
\hat{M}^{(1,3)} = \begin{bmatrix}
8  &  -3 &  1  &  -2  &  0  &  0  &  0  &  0  &  0  &  0  &  0  &  3  &  3\\
1  &  8  & -3  &  -2  &  0  &  0  &  0  &  0  &  0  &  0  &  3  &  0  &  3\\
-3 &  1  &  8  &  -2  &  0  &  0  &  0  &  0  &  0  &  0  &  3  &  3  &  0
\end{bmatrix} \,,\quad\hat{M}^{(2,2)} = \begin{bmatrix}
 6  &   2  &   6  &  0  &  0  &  0\\
 6  &   6  &   2  &  0  &  0  &  0\\
 2  &   6  &   6  &  0  &  0  &  0\\  
-1  &  -5  &  -1  &  0  &  0  &  0\\
-1  &  -1  &  -5  &  0  &  0  &  0\\
-5  &  -1  &  -1  &  0  &  0  &  0\\  
 0  &  -3  &   0  &  0  &  0  &  0\\
 0  &   0  &  -3  &  0  &  0  &  0\\
-3  &   0  &   0  &  0  &  0  &  0
\end{bmatrix} \, .
\eea
The three-loop symbol has 1,773 terms.  One of its sewing matrices is
\bea
\hat{M}^{(3,3)} &=& \scalemath{0.6}{\begin{bmatrix}
-68 & 42 & -10 & 8 & 0 & 12 & -12 & 0 & 0 & 0 & 0 & -18 & -30\\
-10 & -68 & 42 & 8 & -12 & 0 & 12 & 0 & 0 & 0 & -30 & 0 & -18\\ 
42 & -10 & -68 & 8 & 12 & -12 & 0 & 0 & 0 & 0 & -18 & -30 & 0\\
14 & -49 & 12 & 2 & 0 & 0 & -4 & 0 & 0 & 0 & 0 & 0 & -2\\ 
12 & 14 & -49 & 2 & -4 & 0 & 0 & 0 & 0 & 0 & -2 & 0 & 0\\
-49 & 12 & 14 & 2 & 0 & -4 & 0 & 0 & 0 & 0 & 0 & -2 & 0\\ 
-9 & 0 & -10 & 9 & 0 & 0 & 0 & 2 & 0 & 0 & -9 & 0 & 0\\
-10 & -9 & 0 & 9 & 0 & 0 & 0 & 0 & 2 & 0 & 0 & -9 & 0\\ 
0 & -10 & -9 & 9 & 0 & 0 & 0 & 0 & 0 & 2 & 0 & 0 & -9\\
2 & 5 & 5 & -8 & 0 & -2 & 2 & 0 & 0 & 0 & 0 & -1 & 1\\ 
5 & 2 & 5 & -8 & 2 & 0 & -2 & 0 & 0 & 0 & 1 & 0 & -1\\
5 & 5 & 2 & -8 & -2 & 2 & 0 & 0 & 0 & 0 & -1 & 1 & 0\\ 
13 & -16 & -9 & 5 & 0 & 4 & -4 & 4 & 0 & 0 & -5 & 2 & -2\\
-9 & 13 & -16 & 5 & -4 & 0 & 4 & 0 & 4 & 0 & -2 & -5 & 2\\ 
-16 & -9 & 13 & 5 & 4 & -4 & 0 & 0 & 0 & 4 & 2 & -2 & -5\\
5 & -9 & -2 & 4 & 0 & 0 & 0 & 0 & 0 & 0 & 0 & 0 & 0\\ 
-2 & 5 & -9 & 4 & 0 & 0 & 0 & 0 & 0 & 0 & 0 & 0 & 0\\
-9 & -2 & 5 & 4 & 0 & 0 & 0 & 0 & 0 & 0 & 0 & 0 & 0\\ 
10 & -8 & 8 & -1 & 2 & 2 & -2 & 0 & 0 & 0 & 1 & 1 & -1\\
8 & 10 & -8 & -1 & -2 & 2 & 2 & 0 & 0 & 0 & -1 & 1 & 1\\ 
-8 & 8 & 10 & -1 & 2 & -2 & 2 & 0 & 0 & 0 & 1 & -1 & 1
\end{bmatrix}} \,,
\label{M3333} 
\eea
which is considerably more compact.

We provide additional sewing matrices for the $\trphi3$ form factor, through 5 loops, in the ancillary file {\tt phi3SewMatrix.txt}.

\subsection{Resummed OPE results}
\label{sec:resumOPE}

From the final results for $\EE_{3,3}$, it is straightforward to extract the near-collinear limit in a form that is exact in $S$.  We use the same results for the behavior of the functions in $\cC$ through weight 8 in this limit that were used to obtain the resummed OPE limits for the $\trphi2$ form factor~\cite{Dixon:2022rse}.  It is also straightforward to convert the results to the Wilson loop normalization, $\cW_{3,3}$.  We have carried out this exercise for the first two orders, $T^1$ and $T^3$.  

The form of the $T^1$ term in the OPE expansion is,
\be
\cW_{3,3}^{(L)}\ \supseteq\ T \, \frac{S}{1+S^2} \sum_{k=0}^L \cW^1_{L,k}(S) \, \ln^k T \,,
\label{WT1exp}
\ee
where the coefficients $\cW^1_{L,k}(S)$ are weight $2L-k$ functions that are expressible in terms of HPLs, or equivalently the $G$ functions defined in \eqn{eq:G_func_def}, where the indices $a_i \in \{0,-1\}$ and the argument is $S^2$.  The $\cW^1_{L,k}(S)$ are provided through $L=6$ loops in the ancillary file {\tt WL\_FFOPE.txt}.

The form of the $T^3$ term in the OPE expansion is,
\be
\cW_{3,3}^{(L)} \supseteq T^3 \sum_{k=0}^L \biggl[ S\,\cW^{3,A}_{L,k}(S) 
+ \frac{1}{S}\,\cW^{3,B}_{L,k}(S) 
+ \frac{S}{1+S^2}\,\cW^{3,C}_{L,k}(S) 
+ \frac{S\left(2+S^2\right)}{\left(1+S^2\right)^2}\,\cW^{3,D}_{L,k}(S) \biggr] \ln^k T \,.
\label{WT3exp}
\ee
The four coefficients $\cW^{3,X}_{L,k}(S)$, $X = A,B,C,D$, are again $G$ functions with indices $a_i\in\{0,-1\}$. However, because the transcendental functions are now being expanded to subleading power in $T$, they do not have uniform transcendental weight.  They do have a maximum weight of $2L-k$, and this maximum is only achieved for $\cW^{3,D}_{L,k}(S)$.  The coefficient functions $\cW^{3,X}_{L,k}(S)$ are provided through $L=6$ loops in the same ancillary file {\tt WL\_FFOPE.txt}.

\begin{figure}[t]
  \centering
  \begin{minipage}[b]{0.55\textwidth}
    \includegraphics[width=\textwidth]{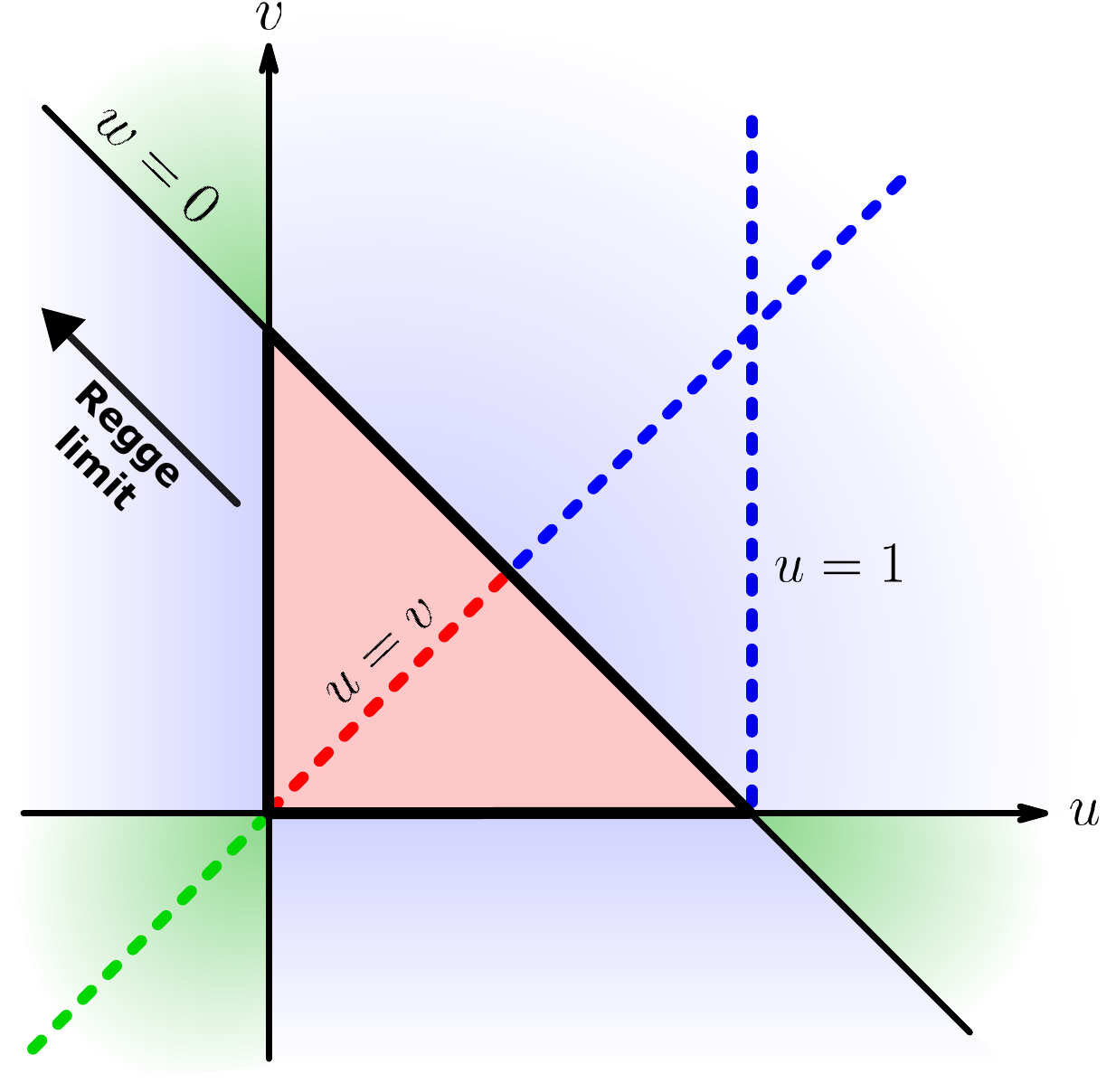}
  \end{minipage}
  \caption{The kinematic space of the three-point form factor. The red triangle represents the Euclidean region. The three lines forming its boundaries correspond to the collinear limits, which are accessible via the FFOPE. The blue and green areas denote the scattering regions, with the blue region corresponding to spacelike operator momentum, and the green region corresponding to timelike operator momentum. To sample the behavior of $\EE_{3,3}$ in each of these regions, we have evaluated it numerically along the dashed lines shown in the figure. Lastly, we have indicated the location of the Regge limit we are considering, where both $u$ and $v$ are sent to infinity, while $w$ is held fixed.}
  \label{Fig:kinematic_regions}
\end{figure}

\subsection{Numerical results}
\label{sec:numerics}

We now describe some of the numerical properties of the $\trphi3$ form factor $\EE_{3,3}$ and its remainder function $R_{3,3}$ for various kinematics. We can identify three distinct regions in the two-dimensional kinematic space of the three-point form factor, which is illustrated in figure~\ref{Fig:kinematic_regions}.
The Euclidean region corresponds to positive values of all three kinematic variables, $0<u,v,w<1$, namely the red triangle in the figure.  If one of these variables becomes negative, while the other two remain positive, we find ourselves in the Minkowski scattering region with spacelike operator momentum (the blue regions). If only one of the three remains positive, then the momentum of the operator is timelike (the green regions).

We start with numerical results at the symmetric Euclidean point where $u=v=w=1/3$.
The form factor is expressible analytically in terms of cyclotomic zeta values involving $6^{\rm th}$ roots of unity~\cite{Ablinger:2011te,HyperlogProcedures,Caron-Huot:2019bsq,Dixon:2021tdw}.
In the $6^{\rm th}$ root of unity $f$-alphabet ``{\tt f23}'' in ref.~\cite{HyperlogProcedures}, there are two weight 1 letters, $f^6_{\pm1}$ and one letter for each higher integer weight, $f^6_{2}$, $f^6_{3}$, $f^6_{4}$, $\ldots$.
In this representation, the first few analytic values of $\EE_{3,3}$ are
\bea
\EE_{3,3}^{(1)}(\tfrac{1}{3},\tfrac{1}{3},\tfrac{1}{3}) &=& -\,3\,\zeta_2 \,, \label{eq:EE33sym1}\\
\EE_{3,3}^{(2)}(\tfrac{1}{3},\tfrac{1}{3},\tfrac{1}{3}) &=&  -\,\frac{351}{32}\,f^6_{3,1}
 + \frac{99}{8}\,\zeta_4 \,, \label{eq:EE33sym2} \\
\EE_{3,3}^{(3)}(\tfrac{1}{3},\tfrac{1}{3},\tfrac{1}{3}) &=&  \frac{31455}{512} \,f^6_{3,3} + \frac{945}{128}\,f^6_{3,-1,-1,1}
- \frac{567}{64}\,f^6_{3,-1,1,1} + \frac{351}{16}\,f^6_{3,1,1,1} \nonumber\\
&&\null - \frac{766989}{20480}\,f^6_{5,1}
+ \zeta_2\,\Bigl( \frac{351}{16}\,f^6_{3,1} + \frac{315}{32}\,f^6_{3,-1} \Bigr)
-\frac{275553}{6656}\,\zeta_6 \,, \label{eq:EE33sym3} 
\eea
or numerically,
\bea
\EE_{3,3}^{(1)}(\tfrac{1}{3},\tfrac{1}{3},\tfrac{1}{3}) &=& -4.934802200544679309417 \,, \label{eq:EE33symnum1}\\
\EE_{3,3}^{(2)}(\tfrac{1}{3},\tfrac{1}{3},\tfrac{1}{3}) &=&  13.7031625477023077256893 \,, \label{eq:EE33symnum2} \\
\EE_{3,3}^{(3)}(\tfrac{1}{3},\tfrac{1}{3},\tfrac{1}{3}) &=&  53.5364253802286696548780 \,. \label{eq:EE33symnum3} 
\eea
%

\begin{table}
\begin{center}
\begin{tabular}{|c|r|c|c|}
\hline\hline
$L$ & $R_{3,3}^{(L)}(\frac{1}{3},\frac{1}{3},\frac{1}{3})$
    & $\frac{R_{3,3}^{(L)}(\frac{1}{3},\frac{1}{3},\frac{1}{3})}
      {R_{3,3}^{(L-1)}(\frac{1}{3},\frac{1}{3},\frac{1}{3})}$ & 
      $\frac{\EE_{3,3}^{(L)}(\frac{1}{3},\frac{1}{3},\frac{1}{3})}
      {\EE_{3,3}^{(L-1)}(\frac{1}{3},\frac{1}{3},\frac{1}{3})}$ \\
\hline\hline
1  &     0                & --            & --\\ 
2  &     $-$14.70782233   & --           & ~$-$2.77684\\
3  &       198.60406266   &  $-$13.50329 & ~~~3.90686 \\
4  &   $-$2668.51585319   &  $-$13.43636 & $-$34.88756 \\
5  &     36649.83756944   &  $-$13.73416 & $-$17.17696 \\
6  & $-$513342.78269312   &  $-$14.00668 & $-$15.27945 \\
\hline\hline
\end{tabular}
\caption{\label{tab:R_symmax} The value of the $L$-loop remainder function $R_{3,3}^{(L)}$ at the Euclidean symmetric point, $(u,v,w)=(\frac{1}{3},\frac{1}{3},\frac{1}{3})$, as well as the ratio to the previous loop order. We also give the successive-loop ratio for $\EE_{3,3}^{(L)}$.}
\end{center}
\end{table}

Table~\ref{tab:R_symmax} gives the values of the remainder function $R_{3,3}^{(L)}$ at the symmetric Euclidean point, through 6 loops.  The table also gives the ratios of successive loop orders for both $R_{3,3}$ and $\EE_{3,3}$.  At large $L$, we expect the successive-loop ratios to approach the same value as for the cusp anomalous dimension, where $\Gcusp^{(L)}/\Gcusp^{(L-1)} \to -16$ at large $L$~\cite{Beisert:2006ez}.  The approach for $R_{3,3}$ seems slower than for the remainder function for the $\trphi2$ at the same point (see Table~8 of ref.~\cite{Dixon:2022rse}).  For $\EE_{3,3}$ at 6 loops the ratio is closer to $-16$ than it is for the $\EE$ functions for $\trphi2$ are at 6 loops; however, the ratio is not yet moving consistently toward $-16$.

The two-loop value for $R_{3,3}$ in Table~\ref{tab:R_symmax} differs from the remainder function in ref.~\cite{Brandhuber:2014ica} by $4\zeta_4$, due to the $g^4$ term in \eqn{eq:C22}. The three-loop value in Table~\ref{tab:R_symmax} differs from the value $160.48 \pm 0.22$ reported in eq.~(5.28) of ref.~\cite{Lin:2021qol} by $-38.12 \pm 0.22$.   This difference is consistent with the numerical value of the $g^6$ term in \eqn{eq:C22}, from refs.~\cite{Guan:2023gsz,Chen:2023hmk}, $-38.260645\ldots$, and our choice of constants, $C_{3,3} = C_{2,2}$, see \eqn{eq:C33isC22}. This numerical agreement with a direct computation provides an indirect test of the FFOPE framework all the way down to the level of constants.

\begin{figure}[t]
  \centering
  \begin{minipage}[b]{0.48\textwidth}
    \includegraphics[width=\textwidth]{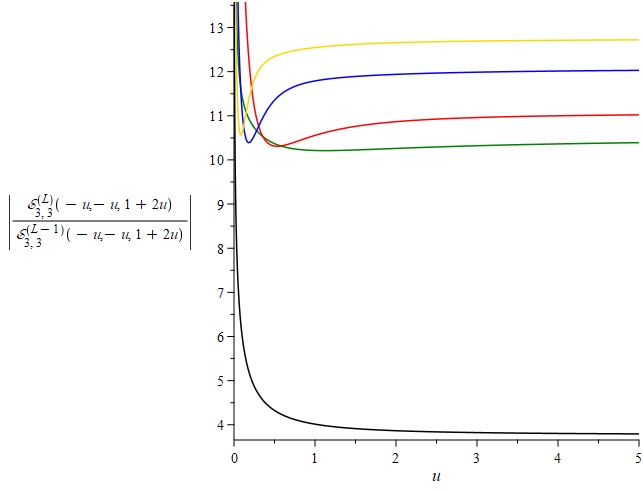}
  \end{minipage}
  \hfill
  \begin{minipage}[b]{0.48\textwidth}
    \includegraphics[width=\textwidth]{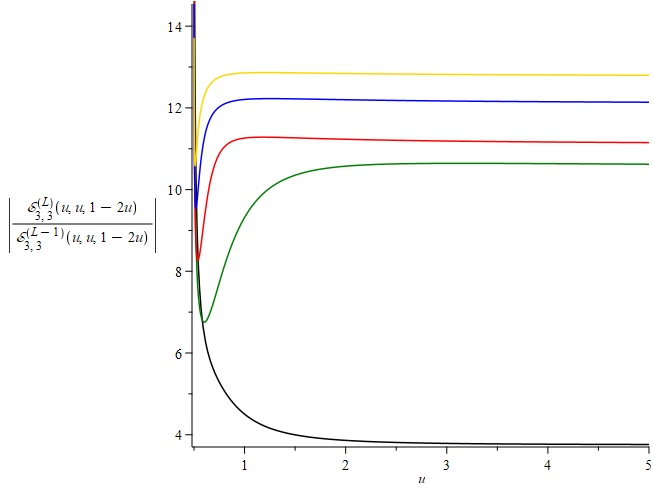}
  \end{minipage}
  \hfill
  \begin{minipage}[b]{0.96\textwidth}
    \includegraphics[width=\textwidth]{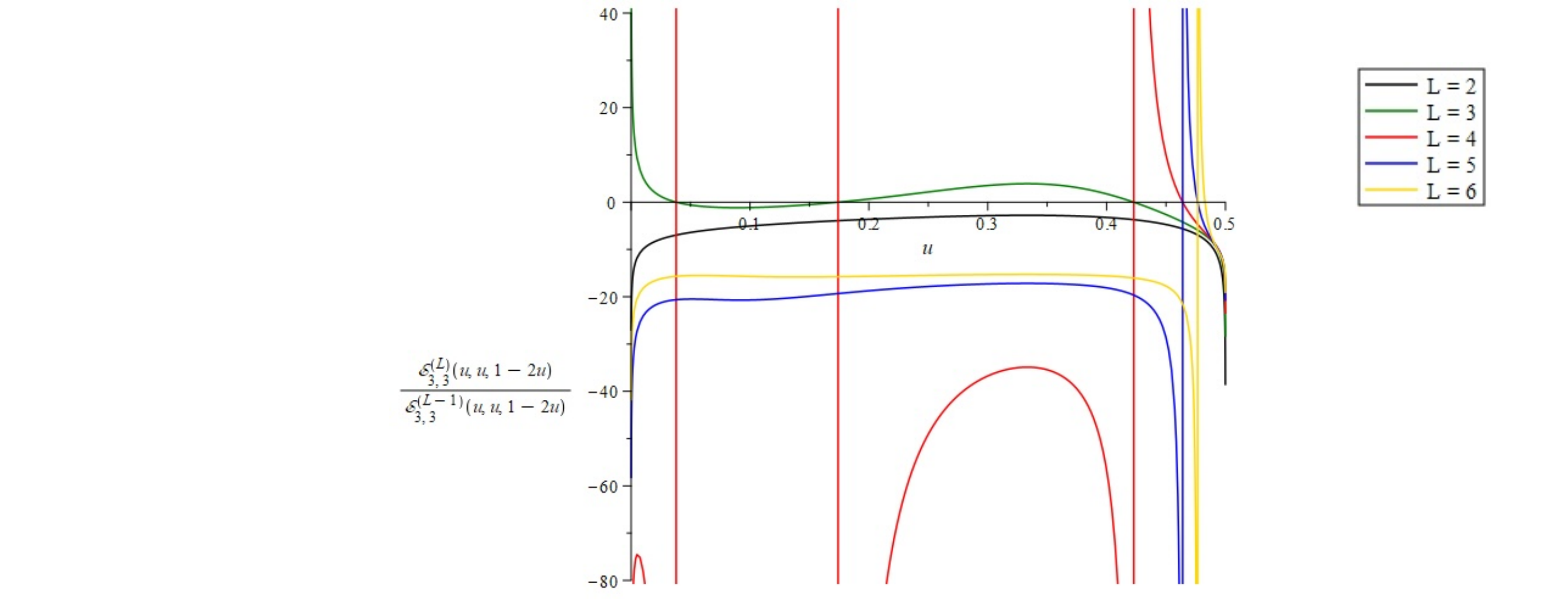}
  \end{minipage}
  \caption{The ratio of $\EE_{3,3}$ at successive loop orders along the symmetric $u = v$ line. Here, the top left plot represents the timelike scattering region, while the top right plot represents the spacelike scattering region. The bottom plot illustrates the behavior in the Euclidean region.}
  \label{Fig:u=v}
\end{figure}

Let us now study the form factor numerically on specific lines in this kinematic space. The symmetric line $\left(u,u,1-2u\right)$, characterized by $u=v$, is especially of interest to us due to its simplicity and the fact that it passes through all three kinematic regions. For $u<0$, we find ourselves in the timelike (operator momentum) scattering region, while $0<u<\frac{1}{2}$ corresponds to the Euclidean region. The segment of the line with $u>\frac{1}{2}$ lies in the spacelike scattering region. The form factor in all regions can be expressed in terms of HPLs with indices $a_i\in\{-1,0,1\}$. It is useful to adopt different HPL arguments in different regions, in order to be able to perform high order series expansions that converge in each region. The numerical behavior of the form factor within these distinct regions is depicted in figure~\ref{Fig:u=v}. In all of the figures, we choose to normalize $\EE_{3,3}$ at the $L$th loop order by its value at the $L-1$ loops, as we did in the third column of Table~\ref{tab:R_symmax}.  In the Euclidean region, $\EE_{3,3}$ is real, but in both scattering regions it has imaginary parts so we plot the absolute magnitude of the ratio.  

\begin{figure}[t]
  \centering
  \begin{minipage}[b]{1.0\textwidth}
    \includegraphics[width=\textwidth]{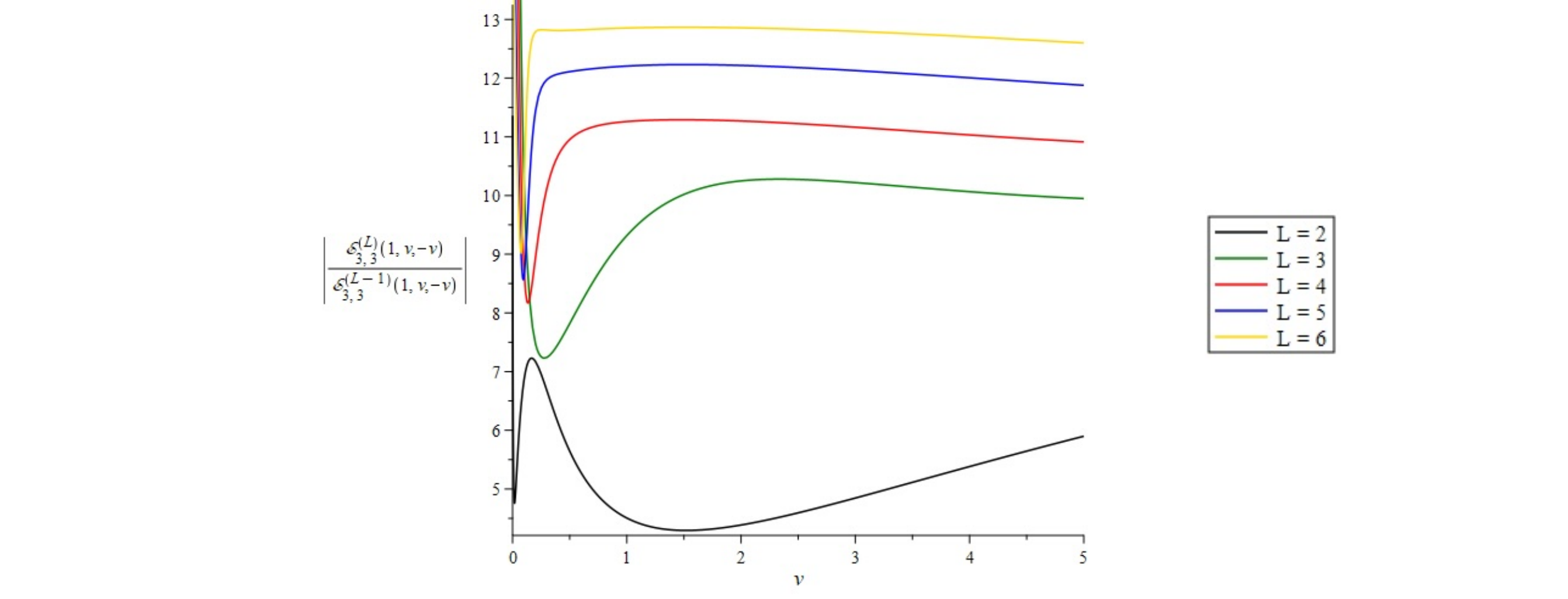}
  \end{minipage}
  \caption{The ratio of $\EE_{3,3}$ at successive loop orders along the $u = 1$ line, with $v>0$.}
  \label{Fig:u=1}
\end{figure}

Another simple line corresponds to setting $u=1$, with positive $v$. It lies entirely within the spacelike scattering region. On this line, the form factor is also expressible in terms of HPLs with indices $a_i\in\{-1,0,1\}$. The behavior of $\EE_{3,3}$ along this line is depicted in figure~\ref{Fig:u=1}.  

\begin{figure}[t]
  \centering
  \begin{minipage}[b]{1.0\textwidth}
    \includegraphics[width=\textwidth]{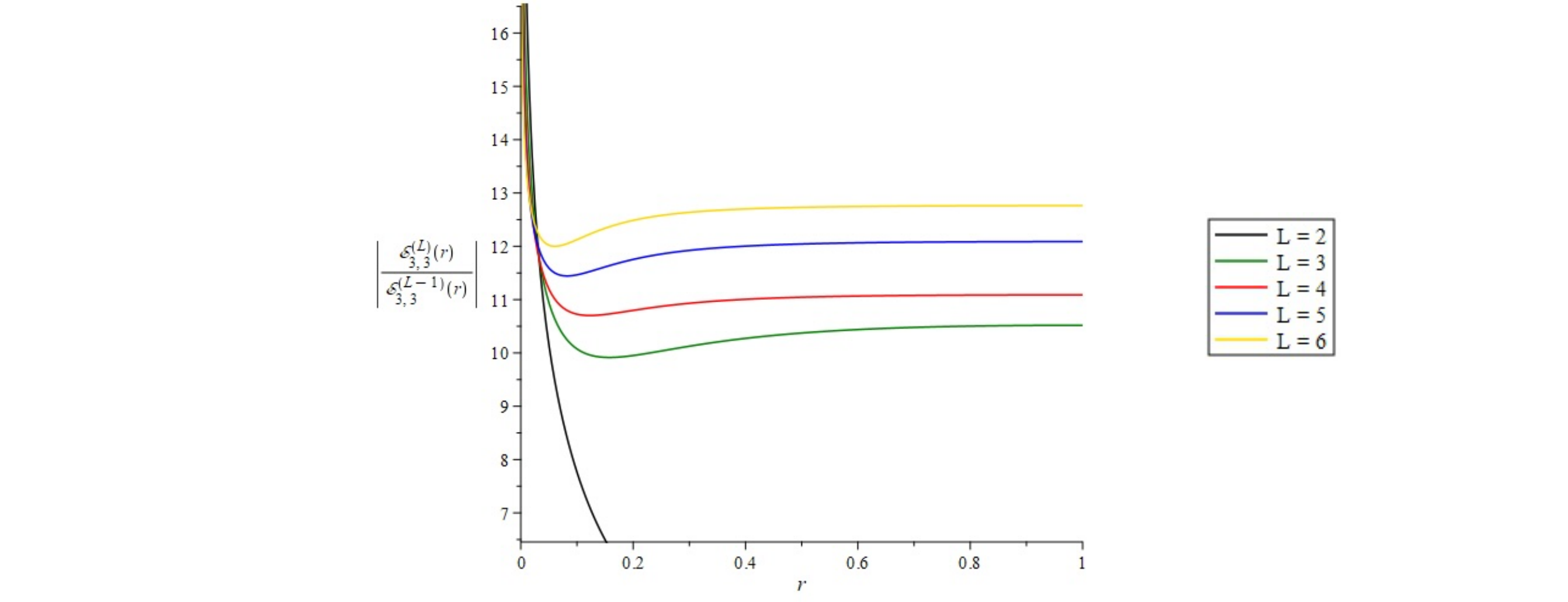}
  \end{minipage}
  \caption{The ratio of $\EE_{3,3}$ at successive loop orders for $u,v\to +\infty$, holding fixed $r = u/v$, i.e.~the fixed-angle, high-energy limit.}
  \label{Fig:fa_r}
\end{figure}

Finally, in figure~\ref{Fig:fa_r} we plot $\EE_{3,3}$ in the spacelike scattering region when $u$ and $v$ are both positive and very large compared to $q^2$, while their ratio is fixed. Since the $q^2$ associated with the operator is negligible, this region can be called fixed-angle high-energy (or lightlike) scattering. In this limit, the $\trphi3$ form factor is finite, but it has nontrivial dependence on the ratio $r=u/v$, which is a measure of the fixed scattering angle. The functions are all HPLs with argument $r$ and indices $a_i\in\{-1,0\}$. In contrast, the three-point form factor of $\trphi2$ has no dependence at all on $r$ in this limit.  The flatness in $r$ for $\trphi2$ can be argued to be due to either a directional dual conformal symmetry~\cite{Lin:2021lqo,Guo:2022qgv} or a final-entry condition on $\EE$~\cite{Dixon:2022rse}.

On all three lines, away from boundary values, the ratios become increasingly flat in the kinematical variable as the loop order increases.  This behavior is consistent with what has been seen for high loop order scattering amplitudes~\cite{Caron-Huot:2019vjl,Dixon:2023kop} and the $\trphi2$ form factor~\cite{Dixon:2022rse}.

\section{Form factor OPE}
\label{sec:FFOPE}

Form factors in planar $\mathcal{N}=4$ SYM admit an integrability description in the form of a near-collinear expansion, called the form factor operator product expansion~\cite{Sever:2020jjx,Sever:2021nsq,Sever:2021xga,Basso:2023bwv}. This approach relies on the existence of a T-dual interpretation of form factors in terms of matrix elements of periodic null Wilson loops, which was first observed at strong coupling in refs.~\cite{Alday:2007he,Maldacena:2010kp}. This dual description was later extended to the weak coupling regime for form factors of $\trphi2$~\cite{Brandhuber:2010ad}, and more recently for form factors of all protected half-BPS operators $\trphi{k}$ in ref.~\cite{Basso:2023bwv}. The dual Wilson loops are naturally described in terms of the dual (region) coordinates $x_i$~\cite{Drummond:2007aua}, with their edges associated with the external momenta of the form factor via $p_i = x_i - x_{i-1}$. Because the local operator carries momentum, the total momentum of the external particles is not conserved, $\sum p_i = q \neq 0$. As a result, the dual Wilson loop is not closed, but periodic.\par
Here, we will explain how the FFOPE results can be obtained at leading order in the near-collinear expansion, but to all orders in the coupling constant. We will then use the FFOPE to predict the Regge limit behavior of the $\trphi3$ form factor. 

\begin{figure}[t]
\centering
\includegraphics[width=.6\textwidth]{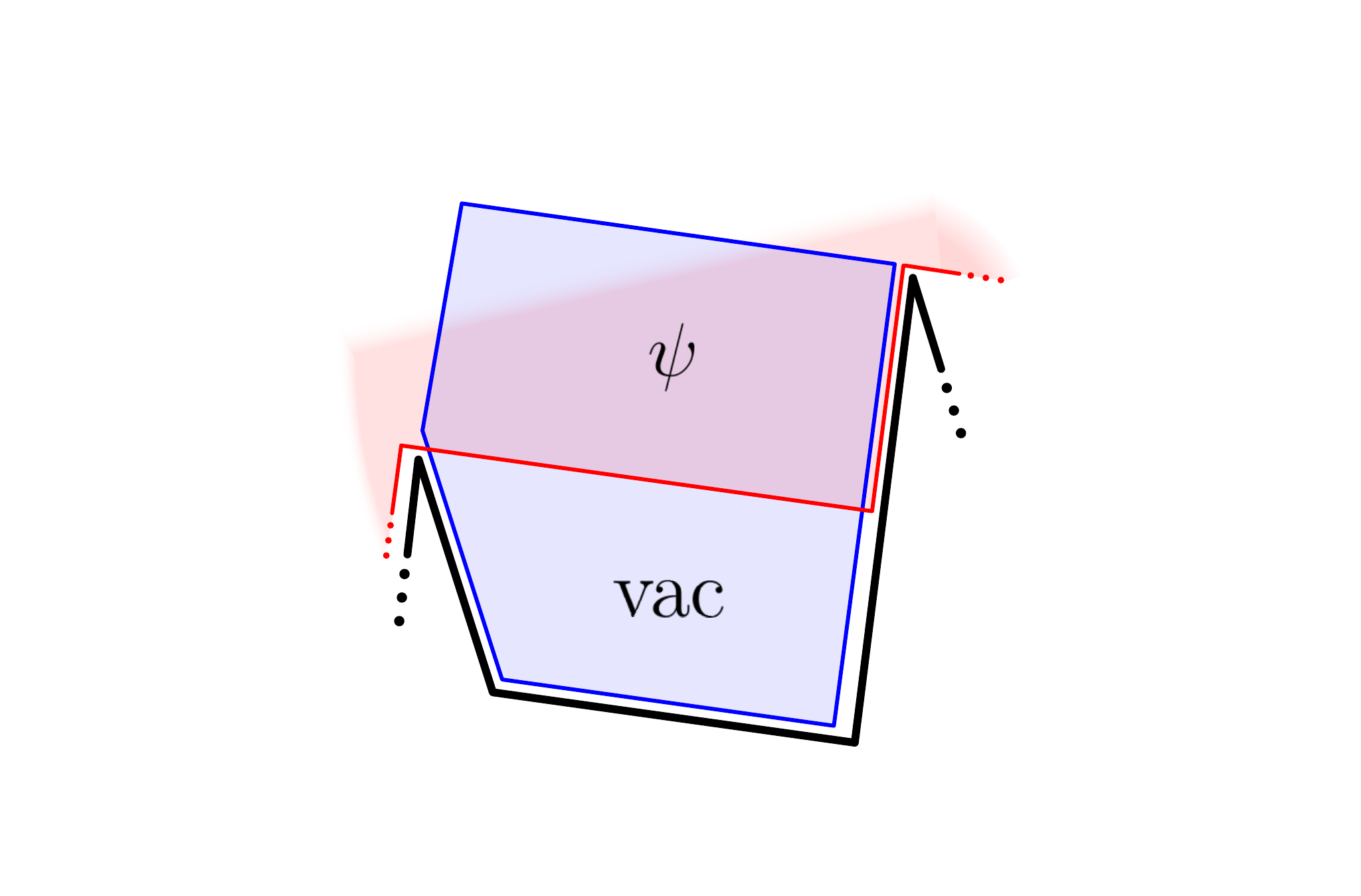}
\caption{Cartoon of a null Wilson loop dual to a three-point form factor. The dots indicate that the contour of the loop (in black) is periodic and repeats itself infinitely. The flux tube dynamics takes place within a single period and is best visualized by splitting the Wilson loop into two null squares, associated with the OPE channels. The bottom channel is filled with the GKP vacuum, while the top channel supports a non-trivial flux tube state $\psi$. Together, the two squares form a null pentagon, associated with the pentagon transition $P(0|\psi)$. The state is absorbed at the top by the form factor transition $F(\psi)$.
}
\label{fig:FFOPE}
\end{figure}

The FFOPE approach recasts the matrix elements of dual Wilson loops as expansions over the Gubser-Klebanov-Polyakov (GKP)~\cite{Gubser:2002tv} flux tube states propagating on top of the string worldsheets bounded by them. When applied to the case of a three-point form factor of the $\trphi3$ operator, it takes the following schematic form, illustrated in figure~\ref{fig:FFOPE},
\begin{equation}\label{FFOPE_def}
\mathcal{W}_{3,3}=\sum\limits_{\psi} e^{-E_\psi\tau+ip_\psi\sigma}\,P(0|\psi)\,F(\psi)\ .
\end{equation}
Here, $\mathcal{W}_{3,3}$ is the specially regularized (framed) Wilson loop that is dual to the $3$-point form factor of $\trphi3$. For precise details on how this regularization is performed, see refs.~\cite{Sever:2020jjx,Basso:2023bwv}. A non-trivial GKP state $\psi$ is created from the vacuum with the use of the pentagon transition $P(0|\psi)$. The absorption of this state by the $\trphi3$ operator is described by the form factor transition $F(\psi)$. The state's propagation through the worldsheet is governed by exponentials of its energy $E_\psi$ and momenta $p_\psi$, which couple to kinematic parameters $\tau$ and $\sigma$. These parameters play the role of the OPE time and space variables and are determined by the external kinematics, as explained in refs.~\cite{Basso:2013vsa,Sever:2020jjx}. The angular OPE variable $\phi$ is manifestly equal to zero in this setup. The flux tube variables $\tau$ and $\sigma$ are related to the conventional $u$, $v$, $w$ parametrization via
\begin{equation}
u = \frac{1}{1+e^{2\sigma}+e^{-2\tau}}\ , \qquad v = \frac{e^{2\sigma}}{\left(1+e^{-2\tau}\right)\left(1+e^{2\sigma}+e^{-2\tau}\right)}\ , \qquad w = \frac{1}{1+e^{2\tau}}\ .
\label{OPEparam3}
\end{equation}
Throughout this paper, we also utilize the variables $S$ and $T$, defined as
\begin{equation}
S=e^\sigma\ ,\quad T=e^{-\tau}\ .
\end{equation}
In this section, however, we will stick to $\sigma$ and $\tau$, as they are more natural from the Wilson loop OPE standpoint.\par
The pentagon transitions $P(\psi|\psi')$, along with the corresponding flux tube measures $\mu(\psi)$, have been studied and bootstrapped at finite coupling in refs.~\cite{Basso:2013vsa,Basso:2013aha,Basso:2014koa,Basso:2014hfa,Basso:2015rta,Basso:2015uxa,Belitsky:2014sla,Belitsky:2014lta,Belitsky:2016vyq}. The form factor transitions for half-BPS operators have likewise been bootstrapped in refs.~\cite{Sever:2021xga,Basso:2023bwv}. As a result, eq.~\eqref{FFOPE_def} provides a fully non-perturbative description of form factors. In general kinematics, this expansion has no small parameters and needs to be resummed. This is, in most cases, an extremely daunting task, which cannot be carried out exactly.%
\footnote{See refs.~\cite{Cordova:2016woh,Lam:2016rel,Belitsky:2017wdo,Bork:2019aud,Drummond:2015jea} for progress in resumming the OPE at weak coupling for scattering amplitudes.}
However, in the collinear limit, which corresponds to $\tau\to\infty$ or $T\to 0$, the expansion \eqref{FFOPE_def} is dominated by the lightest flux tube excitations and doesn't require resummation. We use the exact collinear limit results obtained from it as an extra source of constraints for the perturbative bootstrap.\par
The framed Wilson loop $\mathcal{W}_{3,3}$ has a simple connection to the function $\EE_{3,3}$ introduced in eq.~\eqref{EEtoR},
\begin{equation}\label{W33}
\begin{aligned}
\mathcal{W}_{3,3} &= \Omega_3\,\exp\Big[\Gamma_\text{cusp}\left(\sigma^2+\tau^2+\zeta_2\right)\Big]\,\EE_{3,3} \\
& = \Omega_3\,\exp\Bigg[\frac{\Gamma_\text{cusp}}{4}\Bigg(\ln^2{\left[\frac{w}{1-w}\right]}+\ln^2{\left[\frac{v}{u\,(1-w)}\right]}+4\,\zeta_2\Bigg)\Bigg]\,\EE_{3,3}\ .
\end{aligned}
\end{equation}
The additional normalization factor $\Omega_3$ arises from absorbing the dependence of the Wilson loop on helicity variables, as explained in ref.~\cite{Basso:2023bwv},
\begin{equation}\label{eq:Omega3-data}
\Omega_{3} = \sqrt{uvw} = \frac{1}{\left(e^{\tau}+e^{-\tau}\right)\left(e^{\sigma}+e^{-\sigma}+e^{-2\tau-\sigma}\right)}\ .
\end{equation}
Note that the framed Wilson loop reduces to this factor at tree level, $\mathcal{W}_{3,3} = \Omega_{3} + \mathcal{O}(g^2)$, since $\Gamma_{\textrm{cusp}} = \mathcal{O}(g^2)$ and $\EE_{3,3} = 1+\mathcal{O}(g^2)$ at weak coupling.\par

\subsection{Leading OPE correction}

The leading contribution to the framed Wilson loop~\eqref{W33} comes at the $T^1=e^{-\tau}$ order in the collinear expansion. At tree level, ${\cal O}(g^0)$, it reduces to
\begin{align}\label{W33_tree}
\mathcal{W}_{3,3}^\text{tree} = \Omega_{3} = \frac{e^{-\tau}}{e^\sigma+e^{-\sigma}} + \mathcal{O}(e^{-3\tau})\ .
\end{align}
From the OPE viewpoint, this contribution comes from a state $\psi$ consisting of a single flux tube scalar $\phi$. Both the creation pentagon transition $P(0|\psi)$ and the form factor transition $F(\psi)$ are equal to $1$ in the one-particle case. Hence the entirety of this correction comes from the effective flux tube measure $\tilde{\mu}_{\phi} (u)$ defined in ref.~\cite{Basso:2023bwv},
\begin{equation}\label{eq:W33-data}
\mathcal{W}_{3,3} \overset{e^{-\tau}}{\longrightarrow} \frac{1}{g^{2}}\int \frac{du}{2\pi}\,\tilde{\mu}_{\phi} (u)\ ,\quad\text{where}\quad \tilde{\mu}_{\phi} (u) = \sqrt{\mu_{\phi}(u)\,\nu_{\phi}(u)}\,e^{-E_{\phi}(u) \tau + ip_{\phi}(u)\sigma}\ ,
\end{equation}
The integration variable $u$ is the rapidity of the flux tube excitation, which determines its energy and momentum, 
\begin{equation}\label{Emom}
\begin{aligned}
E_{\phi}(u) &= 1 + 2g^2 \Bigl[ \psi(\tfrac{1}{2}+iu)+\psi(\tfrac{1}{2}-iu)-2\psi(1) \Bigr] + \mathcal{O}(g^4),\\
p_{\phi}(u) &= 2u - 2\pi g^2\,\textrm{tanh}{(\pi u)} + \mathcal{O}(g^4).
\end{aligned}
\end{equation}
The effective measure contains both the regular flux tube measure $\mu_\phi(u)$ and the tilted measure $\nu_\phi(u)$, which describes the form factor. At the leading order in $g^2$, both measures have exactly the same behavior,
\begin{align}
\mu_\phi(u) = \frac{\pi g^2}{\cosh{\pi u}} + \mathcal{O}(g^2) = \nu_\phi(u)\ .
\end{align}
By plugging this formula into eq.~\eqref{eq:W33-data}, along with the leading-order expressions for the energy and momentum, one easily reproduces the tree-level result~\eqref{W33_tree}. At higher loop orders, the two measures start behaving differently. We now briefly recall their all-loop construction.\par
The Wilson loop OPE constituents are constructed with the use of the Beisert-Eden-Staudacher (BES) kernel~\cite{Beisert:2006ez}, which can be represented as a semi-infinite matrix acting in the space of Bessel functions,
\begin{align}\label{eq: BES kernel}
\mathbb{K}_{ij}=2j\left(-1\right)^{ij+j}\int\limits_0^\infty\frac{dt}{t}\,\frac{J_i(2gt)\,J_j(2gt)}{e^t-1}\ .
\end{align}
Here, $i, j = 1, 2, \ldots$ and $J_{i}$ is the $i$-th Bessel function of the first kind. While this kernel is sufficient for describing all amplitude-related OPE constituents, the ones that appear in the form factor analysis require the introduction of a more general object, called the tilted BES kernel~\cite{Basso:2020xts}. It can be cast into the form of a two-by-two block matrix,
\begin{align}\label{eq: tilted BES kernel}
\mathbb{K}(\alpha)=2\cos\alpha\left[\begin{array}{cc}
             \cos\alpha\,\mathbb{K}_{\circ\circ} &\sin\alpha\,\mathbb{K}_{\circ\bullet}\\
            \sin\alpha\,\mathbb{K}_{\bullet\circ} &\cos\alpha\,\mathbb{K}_{\bullet\bullet}
            \end{array}\right]\ ,
\end{align}
where the parameter $\alpha$ is a ``tilt angle" between odd and even Bessel functions, denoted by $\circ$ and $\bullet$, respectively.
The original BES kernel corresponds to $\alpha=\pi/4$. The new type of kernel required for the form factor construction corresponds to $\alpha=0$ and is referred to as the octagon kernel.\par
To construct the OPE measures $\mu_\phi(u)$ and $\nu_\phi(u)$ we need to define the so-called $f$-functions~\cite{Basso:2013pxa,Basso:2013aha,Sever:2021xga,Basso:2023bwv},
\begin{equation}\label{tiltedf}
\begin{aligned}
&f_a^{[\alpha]}(u,v)=\frac{1}{\cos^2{\alpha}}\left[\kappa^u_\alpha\,\mathbb{Q}\,\frac{1}{1+\mathbb{K}(\alpha)}\,\tilde{\kappa}^v_\alpha-\tilde{\kappa}^u_\alpha\,\mathbb{Q}\,\frac{1}{1+\mathbb{K}(\alpha)}\,\kappa^v_\alpha \right] , \\
&f_s^{[\alpha]}(u,v)=\frac{1}{\cos^2{\alpha}}\left[\kappa^u_\alpha\,\mathbb{Q}\,\frac{1}{1+\mathbb{K}(\alpha)}\,\kappa^v_\alpha-\tilde{\kappa}^u_\alpha\,\mathbb{Q}\,\frac{1}{1+\mathbb{K}(\alpha)}\,\tilde{\kappa}^v_\alpha\right] ,
\end{aligned}
\end{equation}
where
\begin{equation}\label{eq:Q}
  \frac{1}{1+\mathbb{K}(\alpha)} = 1- \mathbb{K}(\alpha) + \mathbb{K}(\alpha)^2 -\ldots \ , \quad \mathbb{Q}_{ij} = i \left(-1\right)^{i+1}\delta_{ij}\ .
\end{equation}
The tilted semi-infinite vectors $\kappa^{u}_{\alpha}$ and $\tilde{\kappa}_{\alpha}^{u}$ are defined in block form as
\begin{align}\label{eq: source terms}
\kappa^u_\alpha = 2\cos{\alpha}\left[\begin{array}{cc}
            \sin{\alpha}\,\kappa^u_{\phi,\circ}\\
            \cos{\alpha}\,\kappa^u_{\phi,\bullet}
            \end{array}\right] ,\qquad \tilde{\kappa}^v_\alpha = 2\cos{\alpha}\left[\begin{array}{cc}
            \cos{\alpha}\,\tilde{\kappa}^u_{\phi,\circ}\\
            \sin{\alpha}\,\tilde{\kappa}^u_{\phi,\bullet}
            \end{array}\right] ,
\end{align}
with their $j$-th components given by (for $j = 1,2,\ldots$)
\begin{equation}\label{eq:kappa_scalar}
\begin{aligned}
\kappa_{\phi,j}^u=-\int\limits_0^\infty\frac{dt}{t}\,J_j(2gt)\,\frac{\cos(ut)\,e^{t/2}-J_0(2gt)}{e^t-1}\ ,\,\,\, \tilde{\kappa}_{\phi,j}^u = (-1)^{j+1}\int\limits_0^\infty\frac{dt}{t}\,J_j(2gt)\,\frac{\sin(ut)\,e^{t/2}}{e^t-1}\ .
\end{aligned}
\end{equation}
The function $f_a^{[\alpha]}(u,v)$ is antisymmetric under the argument permutation, while $f_s^{[\alpha]}(u,v)$ is symmetric. Only the former is required for the construction of the tilted OPE measure, which takes the following form,
\begin{equation}\label{tiltedMu}
\mu_\phi^{[\alpha]}(u) = \frac{\pi g^2}{\cosh{\pi u}}\,\exp\left[-\,J_\phi(u)-J_\phi(-u) - f^{[\alpha]}_s(u,u)\right].
\end{equation}
Here, $J_\phi(u)$ is a fixed integral given by
\begin{align}\label{junk}
J_\phi(u) = \frac{1}{2}\int\limits_0^\infty \frac{dt}{t}\left(J_0(2gt)-1\right)\frac{J_0(2gt)+1-2\,e^{t/2-iut}}{e^t-1}\ .
\end{align}
The two measures entering equation~(\ref{eq:W33-data}) correspond to $\alpha=\pi/4$ and $\alpha=0$, respectively,
\begin{equation}
\mu_\phi(u) = \mu_\phi^{[\pi/4]}(u)\ ,\quad \nu_\phi(u) = \mu_\phi^{[0]}(u)\ .
\end{equation}
By systematically expanding the integrand in~\eqref{eq:W33-data} in $g$, one can obtain OPE predictions for the leading $e^{-\tau}$ contribution at any given order in the coupling constant. At every order, the integrand is a meromorphic function in $u$ that can be integrated by closing the contour in the lower half-plane and taking the residues at $u=-\,\frac{i}{2}-in$, with $n=0,1,2,\ldots\,$. These residues are weighted by factors of $e^{(2n+1)\sigma}$, giving the result a natural structure of an expansion around $S=e^\sigma\to 0$. Alternatively, the contour can be closed in the upper half-plane, leading to an expansion around $S\to\infty$. These expansions can then be matched to the perturbative function ansatz, which has the general form discussed in section~\ref{sec:resumOPE}, specifically \eqn{WT1exp}.\par
The subleading corrections to eq.~\eqref{eq:W33-data} arise from dressing the scalar state with singlet pairs of excitations. They appear at the order $\mathcal{O}(e^{-(2m+1)\tau})$ in the collinear expansion, with $m$ being the number of pairs. For these states, the pentagon and form factor transitions are no longer equal to $1$, as they contain non-trivial dynamical and matrix parts. In appendix~\ref{appendix_1}, we discuss the first subleading $e^{-3\tau}$ correction to this equation in detail, which provides an invaluable consistency check for the perturbative bootstrap results.

\subsection{Regge limit}
\label{sec:regge}

In this subsection, we study the behavior of the form factor in the Regge limit, where two Mandelstam invariants tend to infinity. The limit in question is represented by the arrow in figure~\ref{Fig:kinematic_regions}. It corresponds to taking the variables $v, u \rightarrow \infty$, while keeping $w = 1-u-v$ fixed.\footnote{The Regge limit is distinct from the fixed-angle lightlike limit discussed in section~\ref{sec:numerics}, where $u/v$ is held fixed, not $w$.} In this regime, the form factor simplifies drastically and reduces, at each loop order, to a polynomial in the variable
\begin{equation}\label{X}
X\equiv \frac{1}{2}\,\ln\left[\frac{uv}{w^2}\right] = \ln\left[\frac{v}{w}\right] + \frac{i\pi}{2} +\mathcal{O}\left(1/v\right)\, ,
\end{equation}
with $v \approx -\,u \rightarrow \infty -i0$ and $w>0$. For illustration, we find that through six loops, the remainder function adopts the following logarithmic form
\bea
R_{3,3}^{(2)} &=& 2 \, \zeta_2 \, X^2
+ 2 \, \zeta_3 \, X - \frac{19}{4} \zeta_4 \,,
\label{R2Regge}\\
R_{3,3}^{(3)} &=& \frac{8}{3} \, \zeta_3 \, X^3
- 42 \, \zeta_4 \, X^2 - 4 \left( 4 \, \zeta_5 - 3 \, \zeta_2 \zeta_3 \right) X
+ \frac{497}{6}\,\zeta_6 - 4\,\zeta_3^2 \,,
\label{R3Regge}\\
R_{3,3}^{(4)} &=&
4 \, \zeta_4 \, X^4 - 32 \, ( \zeta_5 + \zeta_2 \zeta_3 ) \, X^3
+ ( 701 \, \zeta_6 + 8\,\zeta_3^2 ) \, X^2 \nn\\
&&\null\hskip-0.2cm
+ ( 110 \, \zeta_7 - 168 \, \zeta_2 \zeta_5
  - 388 \, \zeta_4 \zeta_3 ) \, X
- \frac{24551}{24}\,\zeta_8 + 90 \, \zeta_3 \zeta_5
+ 56 \, \zeta_2\zeta_3^2 \,,~~~
\label{R4Regge}\\
R_{3,3}^{(5)} &=& 
\frac{32}{5}\,\zeta_5 \, X^5
- 16 \, ( 12 \, \zeta_6 +\zeta_3^2 ) \, X^4
+ 16 \, ( 30 \, \zeta_7 + 32 \, \zeta_2 \zeta_5
        + 57 \, \zeta_4 \zeta_3  ) \, X^3 \nn\\
&&\null\hskip0cm
- \Bigl( \frac{38606}{3} \, \zeta_8 + 352 \, \zeta_3 \zeta_5
        + 272 \, \zeta_2\zeta_3^2 \Bigr) \, X^2 \nn\\
&&\null\hskip0cm
+ ( -\,168 \, \zeta_9 + 2860 \, \zeta_2 \zeta_7
    + 6168 \, \zeta_4 \zeta_5 + 7802 \, \zeta_6 \zeta_3
    + 56\,\zeta_3^3 ) \, X \nn\\
&&\null\hskip0cm
+ 8393 \, \zeta_{10} - 1196 \, \zeta_3 \zeta_7 - 600\,\zeta_5^2
- 1488 \, \zeta_2 \zeta_3 \zeta_5 - 1214 \, \zeta_4 \zeta_3^2 \,,
\label{R5Regge}\\
R_{3,3}^{(6)} &=& 
\frac{32}{3}\,\zeta_6 \, X^6
- ( 192 \, \zeta_7 + 128 \, \zeta_2 \zeta_5
  + 64 \, \zeta_4 \zeta_3 ) \, X^5 \nn\\
&&\null\hskip0cm
+ \Bigl( \frac{20488}{3} \, \zeta_8 + 608 \, \zeta_3 \zeta_5
  + 320 \, \zeta_2\zeta_3^2 \Bigr) \, X^4 \nn\\
&&\null\hskip0cm
- \Bigl( 8960 \, \zeta_9 + 9600 \, \zeta_2 \zeta_7 + 17664 \, \zeta_4 \zeta_5
   + \frac{61480}{3} \, \zeta_6 \zeta_3
   + \frac{640}{3}\,\zeta_3^3 \Bigl) \, X^3 \nn\\
&&\null\hskip0cm
+ \Bigl( \frac{1350101}{5} \, \zeta_{10} + 7720 \, \zeta_3 \zeta_7
   + 3840 \, \zeta_5^2 + 10464 \, \zeta_2 \zeta_3 \zeta_5
  + 11000 \, \zeta_4 \zeta_3^2 \Bigl) \, X^2 \nn\\
&&\null\hskip0cm
- \Bigl( 21672 \, \zeta_{11} + 57792 \, \zeta_2 \zeta_9 + 115048 \, \zeta_4 \zeta_7 + 142072 \, \zeta_6 \zeta_5
  + \frac{493268}{3} \, \zeta_8 \zeta_3 \nn\\
&&\null\hskip0.3cm
  + 3072 \, \zeta_3^2 \zeta_5 + 1440 \, \zeta_2 \zeta_3^3 \Bigl) \, X \nn\\
&&\null\hskip0cm
+ \frac{331108727}{5528}\,\zeta_{12} + 19656 \, \zeta_3 \zeta_9
+ 18912 \, \zeta_5 \zeta_7 
+ 23500 \, \zeta_2 \zeta_3 \zeta_7 + 11576 \, \zeta_2 \zeta_5^2 \nn\\
&&\null\hskip0cm
+ 40784 \, \zeta_4 \zeta_3 \zeta_5 + 24954 \, \zeta_6 \zeta_3^2
+ 96\,\zeta_3^4 \,.
\label{R6Regge}
\eea
Here, the equalities hold up to power-suppressed corrections in $1/v$.

As alluded to earlier, the logarithmic behavior of the remainder function in eqs.~(\ref{R2Regge})--(\ref{R6Regge}) is similar to the one observed in the multiparticle factorization limit of the six-point NMHV amplitude. This behavior was first studied in perturbation theory through three and four loops in refs.~\cite{Dixon:2014iba,Dixon:2015iva} and extended to all orders in refs.~\cite{BSVfact} using the Pentagon OPE~\cite{Basso:2013aha}. It is also reminiscent of the self-crossing kinematics of the hexagon Wilson loop, analyzed through five loops in ref.~\cite{Dixon:2016epj} and to all orders in ref.~\cite{Caron-Huot:2019vjl} using the all-order Regge formula~\cite{Basso:2014pla}. As we will see below, in line with these studies, the large logarithms of the form factor in the Regge limit may also be summed to all loops using the FFOPE formula.

We can approach the Regge regime starting from the collinear limit, by first taking $w \rightarrow 0$, followed by $v, u \rightarrow \infty$. Despite this limit being different from the general Regge limit described above, the form factor retains the same logarithmic form, with the dependency on $w$ being entirely absorbed in the variable $X$ in eq.~\eqref{X}.
For convenience, we will work in the kinematic regime where $v, u$ are large and imaginary. In terms of the OPE variables, it corresponds to the limits
\begin{equation}\label{eq:OPE-Regge-limits}
\tau \rightarrow \infty\, , \qquad s \equiv -\,2i\,\Big(\sigma +\frac{i\pi}{2}\Big) \rightarrow 0\, ,
\end{equation}
with $s$ real. In these limits, eq.~\eqref{OPEparam3} reduces to
\begin{equation}
u \approx \frac{i}{s}\, , \qquad v \approx -\frac{i}{s}\, , \qquad w \approx e^{-2\tau}\, .
\end{equation}
We also note that in the Regge-OPE limit
\begin{equation}
\tau \approx -\frac{1}{2}\ln{w} \gg 1\, , \qquad -\ln{s} \approx \frac{1}{2}\ln{(uv)} \gg 1\, .
\end{equation}

An important simplification occurs when we take this limit in the FFOPE integral~\eqref{eq:W33-data}. The integral develops a singularity of the form $\sim 1/s$, up to logarithmic corrections. For instance, at tree level, one finds from~\eqref{W33_tree} that the OPE ratio becomes
\begin{equation}
\mathcal{W}_{3,3}^\text{tree} = \sqrt{uvw} = \frac{e^{-\tau}}{s} + \mathcal{O}\left(e^{-3\tau},s^0\right),
\end{equation}
at small $s$ and large $\tau$. One can readily verify that this pole originates from the behavior of the flux-tube measure~\eqref{eq:W33-data} at large rapidity $u\rightarrow \infty$,
\begin{equation}
\tilde{\mu}^{\textrm{LO}}_{\phi}(u) \approx 2\pi g^2 e^{-\tau + 2iu\sigma -\pi u} \,\,\,\, \Rightarrow \,\,\,\, \mathcal{W}_{3,3}^\text{tree} \approx \int\limits_{0}^{\infty} \frac{du}{2\pi g^2}\,  \tilde{\mu}^{\textrm{LO}}_{\phi}(u) = \int\limits_{0}^{\infty} du \, e^{-\tau-u s}\, ,
\end{equation}
where we used $E_{\phi}(u) = 1 + \mathcal{O}(g^2)$ and $p_{\phi}(u) = 2u +\mathcal{O}(g^2)$ at weak coupling.

The formula is easily generalized at higher loops. The key observation is that the measure has a quadratic behavior in $\ln{u}$. At large rapidity, the OPE measure $\mu_{\phi}(u)$, which also governs the six-point NMHV amplitude in the factorization limit, can be parametrized as follows:
\begin{equation}\label{eq:mu-Regge}
\ln{[\mu_{\phi}(u)/\mu^{\textrm{LO}}_{\phi}(u)]} = -\,\frac{\Gamma_{\textrm{cusp}}}{2}\left[\ln{u}+\gamma\right]^2 - \Gamma_{\textrm{virtual}}\left[\ln{u}+\gamma\right] + \Gamma_{\textrm{third}} + \mathcal{O}(1/u)\, ,
\end{equation}
where $\gamma$ is the Euler-Mascheroni constant and $\Gamma_{\textrm{cusp}}$ is the cusp anomalous dimension, given in \eqn{cuspeq}. The next coefficient, $\Gamma_{\textrm{virtual}}$, is the so-called virtual scaling function, governing the subleading behaviour of the scaling dimension of twist-two operators at large spin~\cite{Freyhult:2009my,Fioravanti:2009xt}. It reads
\begin{equation}
\Gamma_{\textrm{virtual}}(g) = -\,12\,\zeta_{3}\,g^{4} + \left(80\,\zeta_{5}+16\,\zeta_{2}\zeta_{3}\right)g^{6} - \left(700\,\zeta_{7} + 80\,\zeta_{2} \zeta_{5} + 168\,\zeta_{4} \zeta_{3}\right) g^{8} + \mathcal{O}(g^{10})\, ,
\end{equation}
through four loops. The third coefficient is less common, but it plays an important role in describing the factorization and self-crossing limits of the hexagon Wilson loops. At weak coupling, it is given by
\begin{equation}
\Gamma_{\textrm{third}}(g) = 5\,\zeta_{2}\,g^{2} - \tfrac{43}{2}\,\zeta_{4}\, g^{4} + \left(\tfrac{925}{6}\,\zeta_{6} - \tfrac{56}{3}\,\zeta_{3}^2\right) g^{6} - \left(\tfrac{5599}{4}\,\zeta_{8} + 12\,\zeta_{2} \zeta_{3}^2 - 260\,\zeta_{3} \zeta_{5}\right) g^{8}+ \mathcal{O}(g^{10})\, .
\end{equation}
All three coefficients may be determined at finite coupling by solving linear equations akin to the BES equation for the cusp anomalous dimension. Namely, one has
\begin{equation}
\Gamma_{\textrm{cusp}} = 4g^2\left[\frac{1}{1+\mathbb{K}}\right]_{11}\, , \qquad \Gamma_{\textrm{virtual}} = 4g\left[\frac{1}{1+\mathbb{K}}\,\kappa^{\textrm{virtual}}\right]_{1}\, ,
\end{equation}
and
\begin{equation}
\Gamma_{\textrm{third}} = \int\limits_{0}^{\infty}\frac{dt}{t} \frac{1-J_{0}(2gt)^2}{e^{t}-1} + \frac{\pi^2}{8}\,\Gamma_{\textrm{cusp}}  - 2\,\kappa^{\textrm{virtual}}\,\mathbb{Q}\,\frac{1}{1+\mathbb{K}}\,\kappa^{\textrm{virtual}}\, ,
\end{equation}
where $\mathbb{K}$ and $\mathbb{Q}$ are the kernels defined in eqs.~\eqref{eq: BES kernel} and~\eqref{eq:Q}, respectively, and where $\kappa^{\textrm{virtual}}$ is a semi-infinite vector, with component $(j\geqslant 1)$
\begin{equation}\label{eq:kappa-virtual}
(\kappa^{\textrm{virtual}})_{j} = \int\limits_{0}^{\infty} \frac{dt}{t} \frac{J_{j}(2gt)\,J_{0}(2gt)-gt\,\delta_{j,1}}{e^{t}-1}\, .
\end{equation}
We obtain a similar expression for the measure $\nu_{\phi}(u)$. In this case, there is no logarithmic behavior, only a constant term (up to exponentially small corrections at large rapidity). We get
\begin{equation}\label{eq:nu-Regge}
\ln{[\nu_{\phi}(u)/\nu^{\textrm{LO}}_{\phi}(u)]} =  \Gamma^{[0]}_{\textrm{third}} + \mathcal{O}(e^{-2\pi u})\, ,
\end{equation}
at large $u$, with
\begin{equation}\label{eq:Gthird-0}
\Gamma^{[0]}_{\textrm{third}} = \int\limits_{0}^{\infty}\frac{dt}{t} \frac{1-J_{0}(2gt)^2}{e^{t}-1} + \frac{\pi^2}{4}\,\Gamma_{\textrm{oct}} - \kappa^{\textrm{virtual}}_{\alpha = 0}\,\mathbb{Q}\,\frac{1}{1+\mathbb{K}(0)}\,\kappa^{\textrm{virtual}}_{\alpha = 0}\, .
\end{equation}
Here,
\begin{equation}
\kappa^{\textrm{virtual}}_{\alpha} = 2\cos{\alpha}\left[\begin{array}{c} \sin{\alpha}\,  \kappa_{\circ}^{\textrm{virtual}} \\ \cos{\alpha}\, \kappa_{\bullet}^{\textrm{virtual}}\end{array}\right]
\end{equation}
is the deformation of the semi-infinite vector $\kappa^{\textrm{virtual}}$ in eq.~\eqref{eq:kappa-virtual}. $\Gamma_{\textrm{oct}}$ is the octagon anomalous dimension, which enters in the description of the origin limits of scattering amplitudes~\cite{Basso:2020xts,Basso:2022ruw} and in the light-like limit of large-charge correlators~\cite{Coronado:2018cxj,Coronado:2018ypq,Kostov:2019stn,Belitsky:2019fan}. It is known explicitly to all loops,
\begin{equation}\label{eq:Gam-oct}
\Gamma_{\textrm{oct}} = 4g^2 \left[\frac{1}{1+\mathbb{K}(0)}\right]_{11} = \frac{2}{\pi^2}\ln{\cosh{(2\pi g)}}\, .
\end{equation}
Expanding the various ingredients in eq.~\eqref{eq:Gthird-0} at weak coupling, one finds that $D\equiv \frac{1}{2}\Gamma^{[0]}_{\textrm{third}}$ admits the representation 
\begin{equation}
D = 4\,\zeta_{2}\,g^2 - 32\,\zeta_{4}\,g^4 + \frac{1024}{3}\,\zeta_{6}\,g^6 - 
 4096\,\zeta_{8}\,g^8 + \frac{262144}{5}\,\zeta_{10}\,g^{10} + \mathcal{O}(g^{12})\, ,
\end{equation}
through five loops. Notice that its coefficients only involve even zeta values, similarly to the weak coupling expansion of $\Gamma_{\textrm{oct}}$ in eq.~\eqref{eq:Gam-oct}. Based on the all-loop formula~(\ref{eq:Gthird-0}), this property must hold to all orders.
Incidentally, this series is identical to the one that gives another exactly known constant $D_{\textrm{oct}}$, showing up in the study of amplitudes and large-charge correlators~\cite{Basso:2020xts,Basso:2022ruw,Coronado:2018cxj,Coronado:2018ypq,Kostov:2019stn,Belitsky:2019fan}.
Namely, we find that $D = D_{\textrm{oct}}$, with~\cite{Basso:2020xts,Belitsky:2019fan}
\begin{equation}
D_{\textrm{oct}} = \ln{\left[\textrm{det} (1+\mathbb{K}(0))\right]} = \frac{1}{4}\ln{\left[\frac{\sinh{(4\pi g)}}{4\pi g}\right]}\, .
\end{equation}
We were unable to prove this relation using the all-order formula~\eqref{eq:Gthird-0}, but we verified that it holds up to high loop orders.

Lastly, we recall the expressions for the energy and momentum of a scalar excitation. In the large rapidity limit, they are also controlled by $\Gamma_{\textrm{cusp}}$ and $\Gamma_{\textrm{virtual}}$. They read~\cite{Basso:2010in,Dorey:2010iy,Dorey:2010id} 
\begin{equation}\label{eq:Ep-large-u}
E_{\phi}(u) = \Gamma_{\textrm{cusp}} \left[\ln{u} +\gamma\right] + 1 + \Gamma_{\textrm{virtual}} + \mathcal{O}(1/u)\, , \qquad p_{\phi}(u) = 2u + \mathcal{O}(1)\, .
\end{equation}

Equipped with the large rapidity behavior of the dispersion relation and the measures, we may proceed to the derivation of an all-loop formula for the Regge limit of the form factor. Plugging eqs.~\eqref{eq:mu-Regge},~\eqref{eq:nu-Regge} and~\eqref{eq:Ep-large-u} into the FFOPE integral, performing a rescaling $u \rightarrow u/s$, and recollecting the logarithms, we find that the BDS-like normalized form factor in \eqn{W33} reads
\begin{equation}\label{eq:E-Regge}
\EE_{3,3} = \int\limits_{0}^{\infty} du \, e^{-u - \frac{1}{4}\Gamma_{\textrm{cusp}} [\ln{u}+\gamma+X]^2 - \frac{1}{2}\Gamma_{\textrm{virtual}} [\ln{u} +\gamma +X] + C}\, ,
\end{equation}
where $X=2\tau - \ln{s}$ in the Regge-OPE regime and
\begin{equation}
\begin{aligned}
C &= \frac{1}{2}\,\Gamma_{\textrm{third}}  -\zeta_{2}\,\Gamma_{\textrm{cusp}} + D \\
& = \tfrac{5}{2}\,\zeta_{2}\,g^2 - \tfrac{91}{4}\,\zeta_{4}\,g^4 + \left(\tfrac{3173}{12}\,\zeta_{6} - \tfrac{28}{3}\,\zeta_{3}^2\right) g^6 + \left(-\,\tfrac{26687}{8}\,\zeta_{8} + 26\,\zeta_{2} \zeta_{3}^2 + 130\,\zeta_{3} \zeta_{5}\right) g^8 + \mathcal{O}(g^{10})\, .
\end{aligned}
\end{equation}
The integral in eq.~\eqref{eq:E-Regge} shares similarities with the one studied in ref.~\cite{Alday:2013cwa} in the light-like limit of correlation functions. Following this reference, and using the integral representation of the Gamma function,
\begin{equation}
\int\limits_{0}^{\infty}du\, e^{-u-z \ln{u}} = \Gamma(1-z) \, ,
\end{equation}
we may cast it into the form
\begin{equation}\label{eq:E-Regge-final}
\EE_{3,3} =  \tilde{\mathcal{E}}_{3,3} \, e^{-\tfrac{1}{4}\Gamma_{\textrm{cusp}} \frac{\partial^2}{\partial z^2}} \left[e^{-z\gamma}\,\Gamma(1-z)\right]\, ,
\end{equation}
with $z \equiv \frac{1}{2}\left[\Gamma_{\textrm{cusp}} X+\Gamma_{\textrm{virtual}}\right]$ and
\begin{equation}
\tilde{\mathcal{E}}_{3,3} \equiv e^{-\tfrac{1}{4}\Gamma_{\textrm{cusp}} X^2 -\tfrac{1}{2}\Gamma_{\textrm{virtual}}X +C}\, .
\end{equation}
The representation~\eqref{eq:E-Regge-final} is quite useful at weak coupling, after replacing the diffusion operator and the Gamma function with their power series,
\begin{equation}
e^{-\tfrac{1}{4}\Gamma_{\textrm{cusp}} \frac{\partial^2}{\partial z^2}} = \sum_{n=0}^{\infty} \frac{1}{n!} \left[-\,\frac{\Gamma_{\textrm{cusp}}}{4}  \,\frac{\partial^2}{\partial z^2}\right]^n\, ,\quad e^{-\gamma z}\,\Gamma(1-z) = \exp{\left[\sum_{n=2}^{\infty} \frac{\zeta_{n}}{n}\,z^{n}\right]}\, .
\end{equation}
At $L$ loops, $\EE_{3,3}$ can be evaluated by truncating both series to the $2L$-th order. Doing so, one may verify the agreement with the 6-loop expressions for the remainder function $R_{3,3}$ in eqs.~(\ref{R2Regge})--(\ref{R6Regge}). In the Regge limit, $\EE_{3,3}^{(1)} = -X^2 +\frac{3\zeta_{2}}{2}$, up to power suppressed corrections, and the relation~\eqref{EEtoR} between the BDS-like normalized form factor and the remainder function becomes
\begin{equation}
\ln{\EE_{3,3}} = R_{3,3} - \frac{\Gamma_{\textrm{cusp}}}{4}\left(X^2 -\frac{3}{2}\,\zeta_{2}\right)\, .
\end{equation}
One may also easily extract predictions at higher loops, by expanding the integral formula~\eqref{eq:E-Regge} to the desired loop order. 

\section{Conclusions}
\label{sec:Conclusions}

In this paper, we presented a calculation of the three-point form factor of the $\trphi3$ operator through six loops in perturbation theory. Our analysis relied on the use of analyticity and integrability methods. They allowed us to build an ansatz for the form factor in terms of MPLs, which was then uniquely fixed using the FFOPE results in the collinear limit. Additional confirmation for our construction is provided by the analysis of the subleading OPE corrections, carried out in appendix~\ref{appendix_1}. We provided the results in a number of ancillary files. We analyzed the numerical behavior of this form factor to high loop orders and derived an all-orders formula for the Regge limit, which matches the perturbative data through six loops.

A remarkable recent observation is that the three-point form factor of $\trphi2$ is antipodally dual to the MHV six-point amplitude in restricted kinematics \cite{Dixon:2021tdw}. There is currently no explanation for this duality, so any kind of generalization would be helpful to shed light on its origins. The three-point form factor of $\trphi3$ studied in this paper can be embedded into the same space of functions $\mathcal{C}$ that supports the antipodal duality in the case of $\trphi2$. As mentioned in section~\ref{sec:c_space}, the function space $\mathcal{C}$ contains a number of adjacency restrictions~(\ref{nonadj}) that do not have a causal explanation, except via the antipodal duality map to the MHV six-point amplitude.
This observation suggests the existence of an antipodal counterpart for the three-point $\trphi3$ form factor. 

A natural candidate for this antipodal counterpart is the six-point NMHV amplitude. However, unlike the three-point form factor of $\trphi3$, the NMHV amplitude is not described by a unique transcendental function of the kinematic variables. It has three independent components in restricted kinematics, which rotate into each other under cyclic transformations.  Only the sum of the three would be dihedrally invariant, as $\trphi3$ is.  Furthermore, it is not easy to identify a kinematic map that would enable such an identification. The main issue is the final entry conditions satisfied by $\EE_{3,3}$, see eqs.~(\ref{FErelationdef}) and~(\ref{FErelationabc}).  They are quite different from the final entry conditions on $\EE$, which map properly to the first entry conditions for MPLs for the six-point amplitude (MHV or NMHV). Hence the final entries for $\EE_{3,3}$ do not align with the first entry relations on the amplitude side, if the kinematic map is to remain the same.

A possible explanation for this negative result is that the three-point antipodal duality is merely a special limit of the more general antipodal self-duality \cite{Dixon:2022xqh}, which relates the four-point MHV form factor of $\trphi2$ to itself. From this perspective, the appearance of the six-point amplitude in the context of the three-point duality might be purely coincidental; it only appears because it happens to describe the triple-collinear limit of the four-point form factor of $\trphi2$. From this perspective, it could be interesting to instead explore generalizations of the four-point antipodal self-duality, to the operator $\trphi3$, for example.

\section*{Acknowledgments}

We thank Dmitry~Chicherin, James~Drummond, {\"O}mer~G{\"u}rdo{\u{g}}an, Andy~Liu, Amit~Sever and Matthias~Wilhelm for interesting discussions. BB and LD thank the Simons Center for Geometry and Physics for hospitality, and LD thanks the Laboratoire de Physique de l'Ecole Normal Sup\'erieure for hospitality, while this work was being completed. AT was supported by the Institut Philippe Meyer at the Ecole Normale Sup\'erieure in Paris.

\appendix
\section{Subleading OPE correction}\label{appendix_1}
While the leading $e^{-\tau}$ order of the collinear expansion happens to be sufficient to fully fix all the remaining coefficients in the perturbative ansatz~\eqref{ansatz} up to six loops, it is useful to have control over the subleading FFOPE contributions as well, as they provide an important consistency check between the integrable and the perturbative descriptions of the form factors. In this appendix, we will describe how the subleading $T^3 = e^{-3\tau}$ correction can be obtained from the FFOPE picture.\par
At tree level, expanding the tree-level framed FFOPE ratio (\ref{eq:Omega3-data}) to the subleading order in $T$ gives us
\begin{align}\label{W33_tree_subleading}
\mathcal{W}_{3,3}^{\textrm{tree}} = \Omega_3 = \frac{e^{-\tau}}{e^\sigma+e^{-\sigma}} - \frac{e^{-3\tau}\left(e^\sigma+2\,e^{-\sigma}\right)}{\left(e^\sigma+e^{-\sigma}\right)^2}+ \mathcal{O}(e^{-5\tau})\ .
\end{align}
The $e^{-3\tau}$ term in this expression is indicative of the presence of a certain set of twist-$3$ excitations contributing to the form factor.
The states of interest have the same R charge as a single scalar $\phi$. They are also required to have a zero $U(1)$ charge, where the group $U(1)$ refers to rotations of the spacetime directions transverse to the flux tube. States of this type can be obtained by dressing a scalar $\phi$ with two-particle singlet states made out of scalars, fermions or gluons, denoted as $\phi\bar{\phi}, \psi\bar{\psi}$ and $F\bar{F}$, respectively.\footnote{One could also consider ``hybrid'' states such as $\psi\psi\bar{F}$ or $\bar\psi\bar\psi F$, which have components in the \textbf{6} of $SU(4)$. However, these states are expected to give rise to vanishing transitions. This statement is a generalization of the conjecture put forward in ref.~\cite{Sever:2021xga} for $\trphi2$ that the multi-particle transitions factorize into products of two- and one-particle transitions. Applied to the form factors of $\trphi3$, this requirement removes all hybrid states, leaving us with states made out of a scalar $\phi$ dressed with singlet pairs at higher twists.}\par
Among these singlets, fermionic excitations behave differently from the other two types, because they produce cuts on the axis of integration in rapidity space. As a result, fermions are traditionally split into two parts living on different Riemann sheets: the large fermion $\psi_L$ and the small fermion $\psi_S$~\cite{Basso:2014koa}. Overall, the total $e^{-3\tau}$ contribution is given by the following sum,\footnote{Unlike what happens with amplitudes~\cite{Basso:2014koa}, there are no mixed fermion terms of the type $\mathcal{W}_{\psi_{L}\phi \bar{\psi}_{S}}$ in the OPE description of form factors of half-BPS operators, because fermions in a pair end up being identified~\cite{Sever:2020jjx}, see eq.~\eqref{eq:deltas} below.}
\begin{equation}
\begin{aligned}
\mathcal{W}_{\bar{X}\phi X} = \mathcal{W}_{\phi}+\mathcal{W}_{\phi\phi\bar{\phi}}+\mathcal{W}_{F\phi\bar{F}}+\mathcal{W}_{\psi_L\phi\bar{\psi}_L}+\mathcal{W}_{\psi_S\phi\bar{\psi}_S} +\mathcal{O}(e^{-5\tau})\ .
\end{aligned}
\end{equation}
Each individual term in this sum is structured as an integral over the product of the pentagon transition that creates the corresponding state and the form factor transition that absorbs it,
\begin{equation}\label{T3}
\begin{aligned}
\mathcal{W}_{\bar{X}\phi X} = \frac{\mathcal{N}_X}{g^{n_X}}\int&\frac{dv_1du\,dv_2}{(2\pi)^3}\,\tilde{\mu}_X(v_1)\,\tilde{\mu}_\phi(u)\,\tilde{\mu}_X(v_2)\,P_{X\phi\bar{X}}(0|v_1,u,v_2)\,F_{X\phi\bar{X}}(v_1,u,v_2)\ .
\end{aligned}
\end{equation}
Here, $\mathcal{N}_X$ is a symmetry factor, equal to $\frac{1}{6}$ for $X=\phi$, and $1$ otherwise. The power $n_X$ corrects the overall coupling scaling and is given explicitly at the end of this appendix. The effective scalar measure $\tilde{\mu}_{\phi}$ was introduced in eq.~(\ref{eq:W33-data}). For $X=\{F,\psi\}$, we define it as
\begin{equation}\label{eq:effective-mu}
\tilde{\mu}_{X}(u) = \mu_{X}(u)\,e^{-E_{X}(u)\tau+ip_{X}(u) \sigma}\ .
\end{equation}
Both the pentagon and form factor transitions appearing in eq.~(\ref{T3}) can be split into their dynamical parts and their coupling-independent matrix parts, which take into account the R-symmetry structures. The dynamical parts of the multi-particle transitions factorize into products of two-particle ones \cite{Basso:2013vsa,Sever:2021xga}. The matrix parts, on the other hand, have to be determined by solving crossing and Watson equations. As we will now demonstrate, however, this can be almost completely avoided for the three-particle transitions.

\subsection{Scalars}
The creation pentagon transition for the scalars has a relatively simple form, which consists of the factorized dynamical part and a matrix part $\Pi$ that multiplies it,
\begin{equation}\label{3pt_pent}
P_{\phi_{i_1}\phi_{i_2}\phi_{i_3}|j}(0|u_1,u_2,u_3)=\frac{1}{P_{\phi\phi}(u_1|u_2)\,P_{\phi\phi}(u_2|u_3)\,P_{\phi\phi}(u_1|u_3)}\,\Pi_{i_1i_2i_3|j}(u_1,u_2,u_3)\ .
\end{equation}
Here, $i_1,i_2,i_3 =1,\ldots, 6$ are the flavor indices of the three scalars, and $j = 1,\ldots , 6$ is a flavor index for the overall representation of the three-scalar state.
The function $P_{\phi\phi}(u|v)$ is the two-particle scalar pentagon transition. As was shown in ref.~\cite{Basso:2023bwv}, it can be unified with another object, which is used in the construction of the form factor transitions. Similarly to the tilted measure~(\ref{tiltedMu}), this object is built out of the $f$-functions~(\ref{tiltedf}),
\begin{equation}\label{tiltedP}
\begin{aligned}
&P_{\phi\phi}^{[\alpha]}(u|v) = \frac{\Gamma\left(iu-iv\right)}{g^2\,\Gamma\left(\frac{1}{2}+iu\right)\Gamma\left(\frac{1}{2}-iv\right)}\,\exp\left[J_\phi(u) + J_\phi(-v) +if^{[\pi/4]}_a(u,v) + f^{[\alpha]}_s(u,v)\right].
\end{aligned}
\end{equation}
The pentagon transition $P_{\phi\phi}(u|v)$ is obtained by setting $\alpha=\frac{\pi}{4}$, while the main building block $Q_{\phi\phi}(u,v)$ of the form factor transitions is obtained by setting $\alpha=0$,
\begin{equation}
P_{\phi\phi}(u|v) = P_{\phi\phi}^{[\pi/4]}(u|v)\ ,\qquad Q_{\phi\phi}(u,v) = P_{\phi\phi}^{[0]}(u|v)\ .
\end{equation}
Similarly to the pentagon transition (\ref{3pt_pent}), the three-scalar form factor transition also consists of a factorized dynamical part and a coupling-independent matrix part. However, as was first observed in ref.~\cite{Sever:2021xga} in the four-scalar case, it contains multiple dynamical structures corresponding to different arrangements of the scalars,
\begin{equation}\label{3pt_FF}
\begin{aligned}
F_{\phi_{i_1}\phi_{i_2}\phi_{i_3}|j}(u_1,u_2,u_3)=&\,\,\frac{Q_{\phi\phi}(u_1,u_2)\,Q_{\phi\phi}(u_2,u_3)}{Q_{\phi\phi}(u_3,u_1)}\,\hat\Pi^{\phi\bar\phi\phi}_{i_1i_2i_3|j}(u_1,u_2,u_3)\\
&+\frac{Q_{\phi\phi}(u_1,u_2)\,Q_{\phi\phi}(u_1,u_3)}{Q_{\phi\phi}(u_3,u_2)}\,\hat\Pi^{\bar\phi\phi\phi}_{i_1i_2i_3|j}(u_1,u_2,u_3)\\
&+\frac{Q_{\phi\phi}(u_1,u_3)\,Q_{\phi\phi}(u_2,u_3)}{Q_{\phi\phi}(u_2,u_1)}\,\hat\Pi^{\phi\phi\bar\phi}_{i_1i_2i_3|j}(u_1,u_2,u_3)\ ,
\end{aligned}
\end{equation}
where we put hats on top of the matrix parts in order to differentiate them from the pentagon transition one, introduced in eq.~(\ref{3pt_pent}).\par
The form factor transition~(\ref{3pt_FF}) satisfies a set of axioms that help us fix these matrix parts. Firstly, the Watson relation states that changing the order of two adjacent scalars results in multiplication by the flux tube S-matrix,
\begin{equation}\label{3pt_Watson}
\begin{aligned}
&S_{\phi\phi}(u_1,u_2)_{i_1\,i_2}^{k_1k_2}\,F_{\phi_{k_2}\phi_{k_1}\phi_{i_3}|j}(u_2,u_1,u_3) = F_{\phi_{i_1}\phi_{i_2}\phi_{i_3}|j}(u_1,u_2,u_3)\ ,\\
&S_{\phi\phi}(u_2,u_3)_{i_2\,i_3}^{k_2k_3}\,F_{\phi_{i_1}\phi_{k_3}\phi_{k_2}|j}(u_1,u_3,u_2) = F_{\phi_{i_1}\phi_{i_2}\phi_{i_3}|j}(u_1,u_2,u_3)\ ,
\end{aligned}
\end{equation}
The S-matrix is given by $S_{\phi\phi}(u,v)_{ij}^{kl} = S_{\phi\phi}(u,v)\,R(u,v)_{ij}^{kl}$, where $S_{\phi\phi}(u,v)$ is the dynamical part computed in refs.~\cite{Basso:2013pxa,Basso:2013aha}, and $R(u,v)_{ij}^{kl}$ is the R-matrix given by
\begin{equation}
R(u,v)_{ij}^{kl} = \frac{u-v}{u-v-i}\,\delta_i^k\delta_j^l -\frac{i}{u-v-i}\,\delta_i^l\delta_j^k + \frac{i\left(u-v\right)}{\left(u-v-i\right)\left(u-v-2i\right)}\,\delta_{ij}\delta^{kl}\ .
\end{equation}
Secondly, the transition~(\ref{3pt_FF}) has to satisfy the crossing axiom,
\begin{equation}\label{3pt_crossing}
F_{\phi_{i_3}\phi_{i_1}\phi_{i_2}|j}(u_3^{2\gamma},u_1,u_2) = F_{\phi_{i_1}\phi_{i_2}\phi_{i_3}|j}(u_1,u_2,u_3)\ .
\end{equation}
Here, the so-called mirror transformation $u\to u^\gamma$ corresponds to the analytic continuation from $u$ to $u+i$. The dynamical parts of the transition transform very non-trivially under this analytic continuation, due to the presence of cuts in the complex rapidity plane. The matrix parts, on the other hand, are rational functions of the rapidities. Hence, the crossing transformation merely acts on them as a shift of the argument: $u^{2\gamma} = u+2i$. Other constraints that can be imposed include the reflection symmetry,
\begin{equation}\label{3pt_reflection}
F_{\phi_{i_3}\phi_{i_2}\phi_{i_1}|j}(-u_3,-u_2,-u_1) = F_{\phi_{i_1}\phi_{i_2}\phi_{i_3}|j}(u_1,u_2,u_3)\ ,
\end{equation}
and the square limit axiom, which fixes the behavior of the transition when a pair of excitations decouples. In our normalization, this axiom reads
\begin{equation}\label{3pt_sq_limit}
\lim\limits_{u'\to u}\left(u-u'\right)F_{\phi_{i_1}\phi_{i_2}\phi_{i_3}|j}(u,v,u') = -\,\frac{2i}{\nu_{\phi}(u)}\,\delta_{i_{1}i_{3}}\delta_{j i_{2}}\ .
\end{equation}
The transformation laws of the dynamical parts are well understood~\cite{Sever:2021xga}, so we can reduce the axioms~(\ref{3pt_Watson}-\ref{3pt_sq_limit}) to algebraic relations for the matrix parts $\hat\Pi^{\bar\phi\phi\phi}$, $\hat\Pi^{\phi\bar\phi\phi}$, $\hat\Pi^{\phi\phi\bar\phi}$. While in general, all three of these structures need to be determined, in practice, it is sufficient to compute just one of them, as the other two are related to it by the Watson axiom. Furthermore, since the creation pentagon transition~(\ref{3pt_pent}) satisfies the same Watson relation, all three structures in eq.~(\ref{3pt_FF}) give identical contributions up to cyclic permutations, after contracting the pentagon and form factor transitions with each other. It is, therefore, sufficient to only focus on $\hat\Pi^{\phi\bar\phi\phi}_{i_1i_2i_3|j}(u_1,u_2,u_3)$, which admits the $SU(4)_{R}$ decomposition
\begin{equation}
\hat\Pi^{\phi\bar\phi\phi}_{i_1i_2i_3|j}(u_1,u_2,u_3) = \pi_1(u_1,u_2,u_3)\,\delta_{ji_1}\delta_{i_2i_3} + \pi_2(u_1,u_2,u_3)\,\delta_{ji_2}\delta_{i_3i_1} + \pi_3(u_1,u_2,u_3)\,\delta_{ji_3}\delta_{i_1i_2}\ .
\end{equation}
The square limit axiom takes the following form for this object,
\begin{equation}
\lim\limits_{u_2\to u_1}\hat\Pi^{\phi\bar\phi\phi}_{i_1i_2i_3|j}(u_1,u_2,u_3) = R (u_1,u_3)_{i_2 i_3}^{i_1 j}\, ,
\end{equation}
where the R-matrix arises from reordering the rapidities using the Watson equation~(\ref{3pt_Watson}). After combining the crossing and Watson axioms and doing some variable redefinitions, one can derive the following set of relations for $\pi_i$,
\begin{equation}\label{3pt_eqns_mp}
\begin{aligned}
\pi_1(u_3,u_2,u_1-2i) &=\frac{u_1-u_2-2i}{\left(u_1-u_2\right)\left(u_2-u_3+i\right)}\left(i\pi_2(u_1,u_2,u_3)+\left(u_2-u_3\right)\pi_3(u_1,u_2,u_3)\right),\\
\pi_2(u_3,u_2,u_1-2i) &=\frac{u_1-u_2-2i}{\left(u_1-u_2\right)\left(u_2-u_3+i\right)}\left(i\pi_3(u_1,u_2,u_3)+\left(u_2-u_3\right)\pi_2(u_1,u_2,u_3)\right),\\
\pi_3(u_3,u_2,u_1-2i) &=\frac{\left(u_1-u_2-2i\right)\left(u_2-u_3-i\right)\left(u_2-u_3-2i\right)}{\left(u_1-u_2\right)\left(u_2-u_3+i\right)\left(u_2-u_3+2i\right)}\,\pi_1(u_1,u_2,u_3)\\
&- \frac{i\left(u_1-u_2-2i\right)\left(u_{2}-u_{3}\right)}{\left(u_1-u_2\right)\left(u_2-u_3+i\right)\left(u_2-u_3+2i\right)}\left(\pi_2(u_1,u_2,u_3)+\pi_3(u_1,u_2,u_3)\right).
\end{aligned}
\end{equation}
Solving these types of equations is usually a fairly complicated task that relies on efficient ansatz construction. Thankfully, in this particular case it can be entirely circumvented! Indeed, it turns out that the solution to the above equations can be easily derived from the matrix part $\hat\Pi^{\phi\bar\phi\phi\bar\phi}_{i_1i_2i_3j}(u_1,u_2,u_3,v)$ of the singlet four-scalar form factor transition constructed in ref.~\cite{Sever:2021xga}.
To do so, we simply need to send the fourth rapidity to infinity and extract the leading-power behavior in $v$.\footnote{Note that the dynamical part of the transitions in ref.~\cite{Sever:2021xga} is expressed in terms of the singlet form factor transition, defined there as
$$F_{\phi\bar\phi}(u,v) = -\,\frac{4}{\left(u-v-i\right)\left(u-v-2i\right)}\,\sqrt{\frac{\nu_\phi(u)\nu_\phi(v)}{\mu_\phi(u)\mu_\phi(v)}}\,Q_{\phi\phi}(u,v)\, ,$$
instead of simply $Q_{\phi\phi}(u,v)$, as is the case in this paper. This results in a slight redefinition of the matrix and dynamical parts.} More precisely, we have
\begin{equation}
\hat\Pi^{\phi\bar\phi\phi}_{i_1i_2i_3|j}(u_1,u_2,u_3) = -\,\frac{1}{4}\lim_{v\to\infty}v^2\,\hat\Pi^{\phi\bar\phi\phi\bar\phi}_{i_1i_2i_3j}(u_1,u_2,u_3,v)\ .
\end{equation}
In terms of $\pi_i$, this gives us
\begin{equation}
\begin{aligned}
\pi_1(u_1,u_2,u_3) &=-\,\frac{2u_1-3u_2+u_3}{\left(u_1-u_2-i\right)\left(u_2-u_3-i\right)\left(u_2-u_3-2i\right)}\ ,\\
\pi_2(u_1,u_2,u_3) &=-\,\frac{1}{\left(u_1-u_2-i\right)\left(u_2-u_3-i\right)}\ ,\\
\pi_3(u_1,u_2,u_3) &=\frac{u_1-3u_2+2u_3}{\left(u_1-u_2-i\right)\left(u_1-u_2-2i\right)\left(u_2-u_3-i\right)}\ .
\end{aligned}
\end{equation}
One can easily verify that this solution satisfies equations~(\ref{3pt_eqns_mp}). Remarkably, the pentagon transition matrix part in eq.~(\ref{3pt_pent}) can be obtained in exactly the same way, using
\begin{equation}
\Pi_{i_1i_2i_3|j}(u_1,u_2,u_3) = \lim_{v\to\infty}v^4\,\Pi_{i_1i_2i_3j}(u_1,u_2,u_3,v)\ .
\end{equation}
This gives us
\begin{equation}
\Pi_{i_1i_2i_3|j}(u_1,u_2,u_3) = \theta_1(u_1,u_2,u_3)\,\delta_{ji_1}\delta_{i_2i_3} + \theta_2(u_1,u_2,u_3)\,\delta_{ji_2}\delta_{i_3i_1} + \theta_3(u_1,u_2,u_3)\,\delta_{ji_3}\delta_{i_1i_2}\ ,
\end{equation}
with
\begin{equation}
\begin{aligned}
\theta_1(u_1,u_2,u_3) &=\frac{u_1-u_3+3i}{\left(u_1-u_2+i\right)\left(u_1-u_3+i\right)\left(u_1-u_3+2i\right)\left(u_2-u_3+i\right)\left(u_2-u_3+2i\right)}\ ,\\
\theta_2(u_1,u_2,u_3) &=-\,\frac{1}{\left(u_1-u_2+i\right)\left(u_1-u_3+i\right)\left(u_1-u_3+2i\right)\left(u_2-u_3+i\right)}\ ,\\
\theta_3(u_1,u_2,u_3) &=\frac{u_1-u_3+3i}{\left(u_1-u_2+i\right)\left(u_1-u_2+2i\right)\left(u_1-u_3+i\right)\left(u_1-u_3+2i\right)\left(u_2-u_3+i\right)}\ .
\end{aligned}
\end{equation}
Contracting the two matrix parts with each other results in
\begin{equation}\label{eq:scalar_MP}
\begin{aligned}
\Pi_{i_1i_2i_3|j}&(u_1,u_2,u_3)\,\hat\Pi^{\phi\bar\phi\phi}_{i_1i_2i_3|k}(u_1,u_2,u_3)\\
&= \frac{6\left(7u_2\left(u_1+u_3-u_2\right)-2u_1^2-3u_1u_3-2u_3^2-4\right)}{\left((u_1-u_2)^2+1\right)\left((u_1-u_2)^2+4\right)\left((u_2-u_3)^2+1\right)\left((u_2-u_3)^2+4\right)}\,\delta_{jk}\\
&\equiv \mathbb{M}(u_1,u_2,u_3) \,\delta_{jk} \ ,
\end{aligned}
\end{equation}
with an implicit sum over repeated indices. As pointed out earlier, the other two structures in eq.~(\ref{3pt_FF}) produce the same expression up to cyclic permutations of the rapidities, resulting in an additional overall factor of $3$, upon integration.

\subsection{Fermions and gluons}
As observed in ref.~\cite{Sever:2020jjx} for $\textrm{Tr} \, \phi^2$, the form factor transitions for singlet pairs of fermions or gluons are localized on the support of delta functions that set the rapidities of conjugated excitations to be equal to each other. We expect the same localization mechanism to be at work for the form factors of $\textrm{Tr} \, \phi^3$. In particular, the form factor transitions for the three-particle states of interest should take the remarkably simple form,
\begin{equation}\label{eq:deltas}
\begin{aligned}
F_{F\phi_i\bar{F}|j}(v_1,u,v_2) &= \frac{2\pi}{\mu_F(v_1)}\,\delta\left(v_1-v_2\right)\delta_{ij},\\
F_{\psi^A\phi_i\bar{\psi}_B|j}(v_1,u,v_2) &= \frac{2\pi}{\mu_{\psi}(v_1)}\,\delta\left(v_1-v_2\right)\delta^A_B\,\delta_{ij}\ ,
\end{aligned}
\end{equation}
where $A, B = 1, \ldots , 4$ are flavor indices for the conjugated fermions.
No matrix parts are present in these expressions, and the second equality is applicable both to small and large fermions. The measures $\mu_{F}$ and $\mu_{\psi}$ controlling the form factors can be found in appendix~\ref{appendix:2pt}.

The pentagon creation transitions for these states are slightly more complicated,
\begin{equation}\label{eq:3pt_GandF}
\begin{aligned}
P_{F\phi_i\bar{F}|j}(0|v_1,u,v_2) &= \frac{1}{P_{F\phi}(v_1|u)\,P_{\phi\bar{F}}(u|v_2)\,P_{F\bar{F}}(v_1|v_2)}\,\delta_{ij},\\
P_{\psi^A\phi_i\bar{\psi}_B|j}(0|v_1,u,v_2) &= \frac{1}{P_{\psi\phi}(v_1|u)\,P_{\phi\bar\psi}(u|v_2)\,P_{\psi\bar{\psi}}(v_1|v_2)}\,\Pi^{A}_{\,\,\,\,\, iB|j}(v_1,u,v_2)\ .
\end{aligned}
\end{equation}
They are made out of two-particle transitions between non-identical excitations $P_{X|Y}$, worked out in refs.~\cite{Belitsky:2014lta,Basso:2015rta}. For the sake of completeness, we recall the expressions for all two-particle transitions in appendix~\ref{appendix:2pt}.\par
In equation~(\ref{eq:3pt_GandF}), the gluon transition has no matrix part. The fermion matrix part may be constructed using the results in ref.~\cite{Belitsky:2016vyq}. It reads
\begin{equation}
\begin{aligned}
\Pi^{A}_{\,\,\,\,\, iB|j}(v_1,u,v_2) = -\,\frac{i\left(u_1-v_1-\frac{3i}{2}\right)\delta_{B}^{A}\delta_{ij}+\frac{i}{2}\left(v_{1}-v_{2}+3i\right)\rho_{i}^{AC}\rho_{jCB}}{\left(u_{1}-v_{1}-\frac{3i}{2}\right)\left(u_{1}-v_{2}+\frac{3i}{2}\right)\left(v_{1}-v_{2}+2i\right)}\, ,
\end{aligned}
\end{equation}
where $\rho_{i}^{AB} = -\rho_{i}^{BA} = (\rho_{i BA})^*$ denote the Weyl components of the 6d Dirac matrices, obeying $\rho_{i}^{AC}\rho_{jCB} + \rho_{j}^{AC}\rho_{iCB} = 2\delta_{ij}\delta_{A}^{B}$. Contracting the pentagon and form factor transitions for the fermion pair, we find
\begin{equation}\label{eq:ferm_matrix}
\delta_{A}^{B}\,\Pi^{A}_{\,\,\,\,\, iB|j}(v,u,v) = -\,\frac{2\left(u-v\right)}{(u-v)^2+\frac{9}{4}}\,\delta_{ij}\ ,
\end{equation}
on the support of the delta function in eq.~\eqref{eq:deltas}.

\subsection{OPE integrals}
We now have all the ingredients needed for evaluating the integrals in eq.~\eqref{T3}. The small fermion contribution happens to dominate in the perturbative limit and constitutes the entirety of the tree-level and one-loop $e^{-3\tau}$ corrections. For this transition, the delta function in eq.~(\ref{eq:deltas}) sets the small fermion rapidities equal, $v_1=v_2=v$. The contour of integration for small fermions~\cite{Basso:2014koa} encircles the singularities located in the lower half of the $v$-plane that arise solely from the matrix part in eq.~(\ref{eq:ferm_matrix}). This fixes $v=u-\frac{3i}{2}$, leaving the integral over the scalar rapidity $u$ as the only non-trivial one. With all the ingredients combined, one finds
\begin{equation}
\begin{aligned}
\mathcal{W}_{\psi_S\phi\bar{\psi}_S} = \frac{i}{g^2}\int\frac{du}{2\pi}\,&e^{-\left(E_\phi(u)+2E_{\psi_S}(u-\frac{3i}{2})\right)\tau + i\left(p_\phi(u)+2p_{\psi_S}(u-\frac{3i}{2})\right)\sigma}\\
&\times\frac{\mu_{\psi_S}(u-\frac{3i}{2})\,\sqrt{\mu_\phi(u)\,\nu_\phi(u)}}{P_{\psi_S\phi}(u-\frac{3i}{2}|u)\,P_{\phi\bar{\psi}_S}(u|u-\frac{3i}{2})\,P_{\psi_S\bar{\psi}_S}(u-\frac{3i}{2}|u-\frac{3i}{2})} \ .
\end{aligned}
\end{equation}
All individual ingredients of this integral can be found in appendix~\ref{appendix:2pt}. This integral can be evaluated order-by-order in the coupling constant by closing the contour in the lower half-plane and taking the residues of the integrand located at $u=-\,\frac{i}{2}-in$, with $n=0,1,2,\ldots\,$. In particular, at tree level, one finds using the formulae in appendix~\ref{appendix:2pt},
\begin{equation}
\mathcal{W}_{\psi_S\phi\bar{\psi}_S}^{\textrm{tree}} = -\,i e^{-3\tau}\int du\,e^{2iu \sigma}\,\frac{u-\frac{3i}{2}}{2\cosh{\, \pi u}}\, ,
\end{equation}
which is in perfect agreement with the term of order $e^{-3\tau}$ in eq.~\eqref{W33_tree_subleading}.\par
At two loops, all four singlet states begin to contribute. The integrands for large fermions and gluons also contain delta functions, but the rapidities of these excitations cannot be uniquely fixed in terms of $u$. Instead, we get two-fold integrals,
\begin{equation}
\begin{aligned}
\mathcal{W}_{\psi_L\phi\bar{\psi}_L} = \frac{1}{g^2}\int\frac{du dv}{(2\pi)^2}\,&e^{-\left(E_\phi(u)+2E_{\psi_L}(v)\right)\tau + i\left(p_\phi(u)+2p_{\psi_L}(v)\right)\sigma}\\
&\times\frac{2\left(v-u\right)}{(u-v)^2+\frac{9}{4}}\,\frac{\mu_{\psi_L}(v)\,\sqrt{\mu_\phi(u)\,\nu_\phi(u)}}{P_{\psi_L\phi}(v|u)\,P_{\phi\bar{\psi}_L}(u|v)\,P_{\psi_L\bar{\psi}_L}(v|v)}\ ,
\end{aligned}
\end{equation}
and
\begin{equation}
\begin{aligned}
\mathcal{W}_{F\phi\bar{F}} = \frac{1}{g^2}\int\frac{du dv}{(2\pi)^2}\,e^{-\left(E_\phi(u)+2E_{F}(v)\right)\tau + i\left(p_\phi(u)+2p_{F}(v)\right)\sigma}\,\frac{\mu_{F}(v)\,\sqrt{\mu_\phi(u)\,\nu_\phi(u)}}{P_{F\phi}(v|u)\,P_{\phi\bar{F}}(u|v)\,P_{F\bar{F}}(v|v)}\ .
\end{aligned}
\end{equation}
The contour of integration is the real line for both scalar and gluon rapidities. For the fermion, one integrates over the real axis, with a $i \varepsilon$ prescription to avoid the cut singularity along the interval $[-2g, 2g]$. (At weak coupling, the cut disappears and is replaced by a pole at $v=0$, whose degree increases with the loop order.)

The three-scalar contribution is the most complicated of the four, as it contains a non-trivial three-fold integral. Explicitly,
\begin{equation}
\begin{aligned}
\mathcal{W}_{\phi\bar{\phi}\phi}& = \frac{1}{2g^6}\int\frac{du_1 du_2 du_3}{(2\pi)^3}\,e^{-\left(E_\phi(u_1)+E_{\phi}(u_2)+E_{\phi}(u_3)\right)\tau + i\left(p_\phi(u_1)+p_{\phi}(u_2)+p_{\phi}(u_3)\right)\sigma}\\
&\times\mathbb{M}(u_1,u_2,u_3)\,\frac{\sqrt{\prod_{i=1}^{3}\mu_{\phi}(u_i)\,\nu_{\phi}(u_i)}}{P_{\phi\phi}(u_1|u_2)\,P_{\phi\phi}(u_2|u_3)\,P_{\phi\phi}(u_1|u_3)}\,\frac{Q_{\phi\phi}(u_1,u_2)\,Q_{\phi\phi}(u_2,u_3)}{Q_{\phi\phi}(u_3,u_1)}\ ,
\end{aligned}
\end{equation}
where $\mathbb{M}(u_1,u_2,u_3)$ is the scalar matrix part defined in eq.~(\ref{eq:scalar_MP}). When evaluating this integral at weak coupling, it is important to note that, in addition to the regular towers of poles at $u_j = -\,\frac{i}{2}-in$, the integrand has poles at $u_2 = u_{1,3}-i$ and $u_2 = u_{1,3}-2i$, which are easy to overlook.\par
Overall, we have produced the $e^{-3\tau}$ FFOPE data for $\trphi3$ up to six loops in perturbation theory, by expanding in $S=e^\sigma$. Higher orders in $S$ correspond to collecting residues from more poles, which becomes computationally expensive at high loop orders.  At six loops, the expansion in $S$ was truncated at $\mathcal{O}(S^{11})$, while at five loops we expanded through $\mathcal{O}(S^{57})$.  In all cases, the bootstrap ansatz was completely fixed before we compared with the $e^{-3\tau}$ terms, and we found perfect agreement with the bootstrap results provided in the ancillary file {\tt WL\_OPE.txt}.

\section{All two-particle pentagon transitions}\label{appendix:2pt}
The two-particle transitions between a pair of twist-one excitations $X$ and $Y$ are given by
\begin{equation}\label{eq:P_XY}
\begin{aligned}
P_{XY}(u|v) =\mathcal{P}_{XY}(u|v)\,\exp\left[J_X(u) + J_Y(-v) +if^{XY}_a(u,v) + f^{XY}_s(u,v)\right] , 
\end{aligned}
\end{equation}
if neither $X$ nor $Y$ is a small fermion. If either of them is a small fermion, we instead have
\begin{equation}\label{eq:P_XY_small}
\begin{aligned}
P_{XY}(u|v) =\mathcal{P}_{XY}(u|v)\,\exp\left[if^{XY}_a(u,v) + f^{XY}_s(u,v)\right] .
\end{aligned}
\end{equation}
The prefactor $\mathcal{P}_{XY}(u|v)$ includes the Born-level transition and some simple higher-loop structures. We provide its expressions for all pairs of excitations below. The exponential phase terms are defined in terms of the $f$-functions,
\begin{equation}
\begin{aligned}
&f_a^{XY}(u,v)=2\left[\kappa^u_X\,\mathbb{Q}\,\frac{1}{1+\mathbb{K}}\,\tilde{\kappa}^v_Y-\tilde{\kappa}^u_X\,\mathbb{Q}\,\frac{1}{1+\mathbb{K}}\,\kappa^v_Y \right] , \\
&f_s^{XY}(u,v)=2\left[\kappa^u_X\,\mathbb{Q}\,\frac{1}{1+\mathbb{K}}\,\kappa^v_Y-\tilde{\kappa}^u_X\,\mathbb{Q}\,\frac{1}{1+\mathbb{K}}\,\tilde{\kappa}^v_Y\right] ,
\end{aligned}
\end{equation}
where $\mathbb{K}$ refers to the original, untilted BES kernel in eq.~(\ref{eq: BES kernel}). The scalar sources $\kappa^u_\phi$ and $\tilde{\kappa}^u_\phi$ are given in eq.~(\ref{eq:kappa_scalar}). The gluon source terms take a similar form,
\begin{equation}
\begin{aligned}
&\kappa_{F,j}^u=-\int\limits_0^\infty\frac{dt}{t}\,J_j(2gt)\,\frac{\cos(ut)\,\exp\left[{(-1)^{j}\,t/2}\right]-J_0(2gt)}{e^t-1}\ ,\\
&\tilde{\kappa}_{F,j}^u = (-1)^{j+1}\int\limits_0^\infty\frac{dt}{t}\,J_j(2gt)\,\frac{\sin(ut)\,\exp\left[{(-1)^{j+1}\,t/2}\right]}{e^t-1}\ .
\end{aligned}
\end{equation}
For the fermions, one finds
\begin{equation}
\begin{aligned}
&\kappa_{\psi_L,j}^u=-\int\limits_0^\infty\frac{dt}{t}\,J_j(2gt)\,\frac{\cos(ut)
-J_0(2gt)}{e^t-1} - \frac{1}{4j}\left(1+(-1)^j\right)\left(\frac{ig}{x(u)}\right)^j\ ,\\
&\tilde{\kappa}_{\psi_L,j}^u = (-1)^{j+1}\int\limits_0^\infty\frac{dt}{t}\,J_j(2gt)\,\frac{\sin(ut)}{e^t-1} - \frac{i}{4j}\left(1-(-1)^j\right)\left(\frac{ig}{x(u)}\right)^j\ ,
\end{aligned}
\end{equation}
and
\begin{equation}
\kappa_{\psi_S,j}^u=\frac{1}{4j}\left(1+(-1)^j\right)\left(\frac{ig}{x(u)}\right)^j\, ,\quad \tilde{\kappa}_{\psi_S,j}^u =  \frac{i}{4j}\left(1-(-1)^j\right)\left(\frac{ig}{x(u)}\right)^j\, .
\end{equation}
Here, $x(u) = \frac{u+\sqrt{u^2-4g^2}}{2}$ is the Zhukowski variable. We also introduce $x^\pm(u) = x\big(u\pm\frac{i}{2}\big)$ for the shifted Zhukowski variables.

The next ingredient in equation~(\ref{eq:P_XY}) comes from the exponential factor $J_X(u)$. Its scalar version was introduced in eq.~(\ref{junk}). For the gluons and large fermions, we have
\begin{equation}\label{junk_F_Psi}
\begin{aligned}
&J_F(u) = \ln\bigg[\frac{x^+(u)\,x^-(u)}{u^2+\frac{1}{4}}\bigg]+\frac{1}{2}\int\limits_0^\infty \frac{dt}{t}\left(J_0(2gt)-1\right)\frac{J_0(2gt)+1-2\,e^{-t/2-iut}}{e^t-1}\ ,\\
&J_{\psi_L}(u) = \ln\bigg[\frac{x(u)}{u}\bigg]+\frac{1}{2}\int\limits_0^\infty \frac{dt}{t}\left(J_0(2gt)-1\right)\frac{J_0(2gt)+1-2\,e^{-iut}}{e^t-1}\ ,
\end{aligned}
\end{equation}
respectively. As mentioned earlier, there is no $J$ factor for the small fermion, $J_{\psi_S}(u) = 0$.

We may now present the full list of prefactors $\mathcal{P}_{XY}$. The ones involving scalars have the following form:
\begin{equation}
\begin{aligned}
& \mathcal{P}_{\phi F}(u|v) = -\,\frac{1}{g}\,\frac{1}{\sqrt{x^+(v)\,x^-(v)}}\,\frac{\Gamma\left(1+iu-iv\right)}{\Gamma\left(\frac{1}{2}+iu\right)\Gamma\left(-\frac{1}{2}-iv\right)}\ ,\quad\mathcal{P}_{\phi \psi_S}(u|v) = \frac{1}{\sqrt{ix(v)}}\ ,\\
&\mathcal{P}_{\phi \psi_L}(u|v) = \frac{1}{g}\,\frac{1}{\sqrt{-ix(v)}}\,\frac{\Gamma\left(\frac{1}{2}+iu-iv\right)}{\Gamma\left(\frac{1}{2}+iu\right)\Gamma\left(-iv\right)}\ ,\quad\mathcal{P}_{\phi\phi}(u|v) = \frac{\Gamma\left(iu-iv\right)}{g^2\,\Gamma\left(\frac{1}{2}+iu\right)\Gamma\left(\frac{1}{2}-iv\right)}\ .
\end{aligned}
\end{equation}
Pure gluonic prefactors are given by
\begin{equation}
\begin{aligned}
&\mathcal{P}_{FF}(u|v) = -\,\frac{1}{g^2}\,Z(u,v)\,\frac{\Gamma\left(iu-iv\right)}{\Gamma\left(-\frac{1}{2}+iu\right)\Gamma\left(-\frac{1}{2}-iv\right)}\ ,\\
&\mathcal{P}_{\bar{F}F}(u|v) = \frac{1}{x^+(u)\,x^+(v)\,x^-(u)\,x^-(v)\,Z(u,v)}\,\frac{\Gamma\left(2+iu-iv\right)}{\Gamma\left(-\frac{1}{2}+iu\right)\Gamma\left(-\frac{1}{2}-iv\right)}\ ,
\end{aligned}
\end{equation}
where
\begin{equation}
Z = \sqrt{1-\frac{g^2}{x^+(u)\,x^+(v)}}\,\sqrt{1-\frac{g^2}{x^+(u)\,x^-(v)}}\,\sqrt{1-\frac{g^2}{x^-(u)\,x^+(v)}}\,\sqrt{1-\frac{g^2}{x^-(u)\,x^-(v)}}\ .
\end{equation}
Transition prefactors involving gluons and fermions are given by
\begin{equation}
\begin{aligned}
&\mathcal{P}_{F\psi_L}(u|v) = -\,\frac{1}{g^2}\,\sqrt{1-\frac{g^2}{x^+(u)\,x(v)}}\,\sqrt{1-\frac{g^2}{x^-(u)\,x(v)}}\,\frac{\Gamma\left(\frac{1}{2}+iu-iv\right)}{\Gamma\left(-\frac{1}{2}+iu\right)\Gamma\left(-iv\right)}\ ,\\
&\mathcal{P}_{F\bar{\psi}_L}(u|v) = -\,\frac{i}{x^+(u)\,x^-(u)\,x(v)\,\sqrt{1-\frac{g^2}{x^+(u)\,x(v)}}\,\sqrt{1-\frac{g^2}{x^-(u)\,x(v)}}}\,\frac{\Gamma\left(\frac{3}{2}+iu-iv\right)}{\Gamma\left(-\frac{1}{2}+iu\right)\Gamma\left(-iv\right)}\ ,
\end{aligned}
\end{equation}
and
\begin{equation}
\begin{aligned}
&\mathcal{P}_{F\psi_S}(u|v) = -\,\frac{u-v+\frac{i}{2}}{x(v)\,\sqrt{1-\frac{g^2}{x^+(u)\,x(v)}}\,\sqrt{1-\frac{g^2}{x^-(u)\,x(v)}}}\ ,\\
&\mathcal{P}_{F\bar{\psi}_S}(u|v) = \sqrt{1-\frac{g^2}{x^+(u)\,x(v)}}\,\sqrt{1-\frac{g^2}{x^-(u)\,x(v)}}\ .
\end{aligned}
\end{equation}
Large fermion transition prefactors are:
\begin{equation}
\begin{aligned}
&\mathcal{P}_{\psi_L\psi_L}(u|v) = \frac{1}{g^2}\,\sqrt{1-\frac{g^2}{x(u)\,x(v)}}\,\frac{\Gamma\left(iu-iv\right)}{\Gamma\left(iu\right)\Gamma\left(-iv\right)}\ ,\\
&\mathcal{P}_{\psi_L\bar{\psi}_L}(u|v) = \frac{1}{x(u)\,x(v)\,\sqrt{1-\frac{g^2}{x(u)\,x(v)}}}\,\frac{\Gamma\left(1+iu-iv\right)}{\Gamma\left(iu\right)\Gamma\left(-iv\right)}\ .
\end{aligned}
\end{equation}
Similarly, for small fermions one finds:
\begin{equation}
\begin{aligned}
\mathcal{P}_{\psi_S\psi_S}(u|v) &= \frac{i\,\sqrt{1-\frac{g^2}{x(u)\,x(v)}}}{u-v}\ ,\qquad \qquad \,\, \mathcal{P}_{\psi_S\bar{\psi}_S}(u|v) = \frac{1}{\sqrt{1-\frac{g^2}{x(u)\,x(v)}}}\ ,\\
\mathcal{P}_{\psi_L\psi_S}(u|v) &= -\,\frac{i}{x(v)\sqrt{1-\frac{g^2}{x(u)\,x(v)}}}\ ,\qquad \mathcal{P}_{\psi_L\bar{\psi}_S}(u|v) = \sqrt{1-\frac{g^2}{x(u)\,x(v)}}\ .
\end{aligned}
\end{equation}
All other transitions can be derived using the following relations:
\begin{equation}
P_{XY}(u|v) = P_{\bar{X}\bar{Y}}(u|v) = P_{YX}(-v|-u) = P_{\bar{Y}\bar{X}}(-v|-u)\ ,
\end{equation}
with $\bar{\phi} = \phi$ for scalar and $\bar{\bar{X}} = X$ for other excitations.

Another important ingredient for the OPE construction is the flux tube measure $\mu_{X}$. These objects can be obtained as residues of pentagon transitions,
\begin{equation}
\underset{v\to u}{\rm res}\,P_{XX}(u|v) = \frac{i}{\mu_X(u)}\ .
\end{equation}
Explicitly,
\begin{equation}
\mu_X(u) = \mathcal{M}_X(u)\,\exp\left[-\,J_X(u)-J_X(-u) - f^{XX}_s(u,u)\right],
\end{equation}
with
\begin{equation}
\begin{aligned}
\mathcal{M}_F(u) &= -\,\frac{1}{Z(u,u)}\,\frac{\pi g^2}{\left(u^2+\frac{1}{4}\right)\cosh{\pi u}}\ , \quad \mathcal{M}_{\phi}(u) = \frac{\pi\,g^2}{\cosh{\pi u}}\ ,\\
\mathcal{M}_{\psi_L}(u) &= \frac{1}{\sqrt{1-\frac{g^2}{x(u)^2}}}\,\frac{\pi\,g^2}{u\sinh{\pi u}}\ ,\quad\mathcal{M}_{\psi_S}(u) = -\,\frac{1}{\sqrt{1-\frac{g^2}{x(u)^2}}}\ .
\end{aligned}
\end{equation}
Lastly, the flux tube energies and momenta of the corresponding excitations are given by:
\begin{equation}
E_X(u) = 1 + 4g\left[\frac{1}{1+\mathbb{K}}\,\kappa^u_X\right]_1\ , \qquad p_X(u) =2u\left(1-\delta_{X\psi_S}\right) - 4g\left[\frac{1}{1+\mathbb{K}}\,\tilde{\kappa}^u_X\right]_1\, ,
\end{equation}
with $\delta_{X\psi_S} = 1$ if $X$ is a small fermion and $\delta_{X\psi_S} = 0$ otherwise.

\bibliographystyle{JHEP}
\bibliography{phi3}

\end{document}